\documentclass[fleqn,usenatbib,useAMS]{mnras}

\usepackage{graphicx}	
\usepackage{amsmath}	
\usepackage{amssymb}	
\usepackage{multicol}        
\usepackage{bm}		
\usepackage{pdflscape}	
\usepackage{nccmath}

\usepackage[T1]{fontenc}
\usepackage{ae,aecompl}

\usepackage{newtxtext,newtxmath}
\usepackage{units}
\usepackage{tablefootnote}
\usepackage[normalem]{ulem}

\title[Pseudomode Frequency Shifts]{A Multi-Instrument Investigation of the Frequency Stability of Oscillations Above the Acoustic Cut-Off Frequency with Solar Activity}

\author[]{K. Kosak$^{1}$\thanks{Contact e-mail: \href{mailto:katie.kosak@gmail.com}{katie.kosak@gmail.com}} R. Kiefer$^{1, 2}$, A.-M. Broomhall$^{1}$
\\
$^{1}$ CFSA, Physics Department, University of Warwick,Coventry CV4 7AL, UK\\
$^{2}$ Leibniz-Institut f\"ur Sonnenphysik (KIS), Sch\"oneckstra\ss e 6, 79104, Freiburg, Germany}
\date{Last updated 2015 May 22; in original form 2013 September 5}

\pubyear{2021}

\hypersetup{draft}
\begin{document}
\label{firstpage}
\pagerange{\pageref{firstpage}--\pageref{lastpage}}
\maketitle

\begin{abstract}
Below the acoustic cut-off frequency, oscillations are trapped within the solar interior and become resonant. However, signatures of oscillations persist above the acoustic cut-off frequency, and these travelling waves are known as pseudomodes. Acoustic oscillation frequencies are known to be correlated with the solar cycle, but the pseudomode frequencies are predicted to vary in anti-phase. We have studied the variation in pseudomode frequencies with time systematically through the solar cycle. We analyzed Sun-as-a-star data from Variability of Solar Irradiance and Gravity Oscillations (VIRGO), and Global Oscillations at Low Frequencies (GOLF), as well as the decomposed data from Global Oscillation Network (GONG) for harmonic degrees $0\le l \le 200$. The data cover over two solar cycles (1996--2021, depending on instrument). We split them into overlapping 100-day long segments and focused on two frequency ranges, namely $5600$--$6800\,\rm\mu Hz$ and $5600$--$7800\,\rm\mu Hz$. The frequency shifts between segments were then obtained by fitting the cross-correlation function between the segments' periodograms. For VIRGO and GOLF, we found no significant variation of pseudomode frequencies with solar activity. However, in agreement with previous studies, we found that the pseudomode frequency variations are in anti-phase with the solar cycle for GONG data. Furthermore, the pseudomode frequency shifts showed a double-peak feature at their maximum, which corresponds to solar activity minimum, and is not seen in solar activity proxies. An, as yet unexplained, pseudo-periodicity in the amplitude of the variation with harmonic degree $l$ is also observed in the GONG data.
\end{abstract}




\section{Introduction}
Helioseismology aims to probe the interior of the Sun by studying the behaviour of the acoustic oscillations, more specifically p-modes. Parameters of the acoustic oscillations, such as frequencies, amplitudes, and lifetimes, vary with the solar cycle, but the underlying mechanisms for these variations are still awaiting precise quantification \citep[see e.g.][and references therein]{2014SSRv..186..191B}. At (or near) the solar surface, p modes are reflected because of the sharp drop in density. The radial positions at which they are reflected depends on frequency, with higher-frequency modes being reflected at shallower depths than lower-frequency modes. However, there is an upper limit to the frequencies at which modes are reflected into the interior, known as the acoustic cut-off frequency (around $5000\,\rm\mu Hz$, e.g. \citealt{2011ApJ...743...99J}). The acoustic oscillations have been vastly studied in the case below the acoustic cut-off. However, studies of oscillations above the acoustic cut-off are far more limited in number. 

Acoustic oscillations above the acoustic cut-off propagate from the photosphere to the base of the solar corona as travelling waves. One would therefore expect a smoothly varying power spectrum above the acoustic cut-off frequency \citep{1991ApJ...375L..35K}. Nevertheless, peaks are observed above the acoustic cut-off frequency in power spectra made from helioseismic data \citep[e.g.][]{1988ESASP.286..279J, 1988ApJ...334..510L, 1991ApJ...373..308D, 1992A&A...266..532F, 1998ApJ...504L..51G, 2003ESASP.517..247C, 2005ApJ...623.1215J}. While oscillations above the acoustic cut-off frequency are not reflected at the solar surface, they are still refracted in the solar interior, just as p modes are. \citet{1990LNP...367...87K} proposed that these high-frequency peaks are caused by interference between modes that initially travel inward from their excitation point, and are refracted in the solar interior, and modes that emanate outwards from their location of excitation. In the past, frequencies above the acoustic cut-off have been referred to as "high-frequency interference peaks" (HIPs) and "pseudomodes". In our study, we use the term "pseudomodes".
 
The in-depth investigation of the pseudomodes holds potential as an additional tool to understand the solar cycle. One example of how pseudomodes can contribute is by defining the precise acoustic cut-off frequency, which is inversely proportional to the density scale height. In the past, the acoustic cut-off was identified by seeing a drop-off of the p-mode frequencies \citep{2006ApJ...646.1398J}. \citet{2006ApJ...646.1398J} proposed using frequency shifts and bivariate analysis of the full frequency spread of p-modes and pseudomodes to find the acoustic frequency to be around $5100\,\rm\mu Hz$. This study was based upon unresolved observations from the Global Oscillations at Low Frequencies \citep[GOLF;][]{1995SoPh..162...61G} and Variability of Irradiance and Gravity Oscillations \citep[VIRGO;][]{1995SoPh..162..101F} instruments, onboard the Solar and Heliospheric Observatory (SoHO). Also using VIRGO data, \citet{2011ApJ...743...99J} found that the acoustic cut-off frequency varies in phase with the solar cycle, changing by around $100$--$150\,\rm \mu Hz$ between cycle minimum and maximum. 

Solar cycle variations in the pseudomode frequencies themselves have also been previously detected. \citet{1994SoPh..150..389R} observed a large (${\sim}15\,\rm\mu Hz$) negative shift in pseudomode frequency between 1991 (a time of high activity) and 1988 (a time of low activity). However, we note that the authors found negative frequency shifts between activity maximum and minimum at around $4000\,\rm\mu Hz$ as well, contrary to other observations \citep[e.g.][]{1990Natur.345..779L}. Using VIRGO data, \citet{2005ApJ...623.1215J} found that the separation of pseudomodes did not vary with the solar cycle. However, \cite{2009AIPC.1170..566S} used GOLF and VIRGO observations for solar cycle 23 to study the amplitude variation of pseudomodes and showed the potential of pseudomodes to investigate the solar cycle. \citet{2011JPhCS.271a2029R} performed a linear regression analysis between acoustic mode frequency and various measures of solar activity, for oscillations both above and below the acoustic cut-off frequency. In agreement with previous studies, \citet{2011JPhCS.271a2029R} demonstrated that below $5000\,\rm\mu Hz$ p-mode frequencies are correlated with solar activity. However, at some point between $5000$ and $5700\,\rm\mu Hz$, there is a switch whereby frequencies become anti-correlated with solar activity. Furthermore, at even higher frequencies, the correlation between mode frequencies switches to being positive again. Additionally, the exact frequency at which these switches from positive to negative correlation and back again happen appears to depend on the phase of the solar cycle under consideration. 

The observed switch in the parity of the correlation is predicted phenomenologically by \citet{1998MNRAS.298..464V}, who modelled the variation in depth of an acoustic potential, which governs the vertical propagation of the waves. \citeauthor{1998MNRAS.298..464V} simulate solar cycle variation by changing the height of the acoustic potential, which impacts the acoustic reflectivity of the solar atmosphere. Alternative models also predict pseudomode shifts that are anti-correlated with the solar cycle and reach a minimum at around $6000\,\rm\mu Hz$, including that of \citet{1995ASPC...76..264J, 1996ApJ...456..399J}, who include magnetic fields in the chromosphere and model the solar cycle in terms of changes in chromospheric temperature (by more than $1000\,\rm K$) and magnetic field strength (by $5\,\rm G$ for \citeauthor{1996ApJ...456..399J} and $20\,\rm G$ for \citeauthor{1995ASPC...76..264J}). 

Our motivation is to study the frequency shift of the pseudomodes as a function of time compared to the solar cycle. Although, as detailed above, comparisons between cycle minimum and maximum have been performed previously, to the best of our knowledge, a systematic study of variations as a function of time has not previously been performed. In this study, we will compare the frequency shifts obtained from different instruments/networks with a proxy for solar activity. The data we used are described in Section \ref{sect:data}. To determine the frequency shift of the pseudomodes we employ a cross-correlation technique, which is described in Section \ref{sect:method}. The results of the study are given in Section \ref{sect:results}, where we consider both unresolved (low harmonic degree) and resolved (intermediate harmonic degree) data separately. Finally, we finish with some concluding remarks in Section \ref{sect:conclusions}.

\section{Data}\label{sect:data}
The VIRGO instrument, on SoHO, comprises three Sun photometers (SPMs) for measuring the solar spectral irradiance at three independent wavelengths: $402\,\rm nm$ (blue), $500\,\rm nm$ (green), and $862\,\rm nm$ (red). Following its launch in December 1996, SoHO was lost for several months, starting in June 1998. SoHO resumed operations in October 1998 with the VIRGO/SPM data of the same high quality as before the "SoHO vacation". Data were produced as described in \citet{1995SoPh..162..101F, 1997SoPh..170....1F, 2002SoPh..209..247J}. This study uses the freely available data, (see Data Availability and footnotes to Table~\ref{tab:1} for URLs), which extend from 1 April 1996 to 30 March 2014 and have a temporal cadence of $60\,\rm s$. 

GOLF is a resonant scatter spectrophotometer, which is also aboard SoHO. It measured line-of-sight velocities using the sodium doublet. GOLF started observing in January 1996, but due to malfunctions, continuous observations began not until April 1996. Before SoHO's vacation, the blue wing of the sodium line was used for observation, but this was changed to the red wing after the vacation. It was later switched back to the blue wing mode of operation on 18 November 2002. To correct for the ageing of GOLF's two detectors (PM1 and PM2), the voltage applied to them was adjusted several times. Table~\ref{tab:A1} gives an overview of the changes made to the GOLF instrument. GOLF observations have a cadence of $20\,\rm s$. Data are processed in the manner described in \citet{2005A&A...442..385G} and recently have been recalibrated by \citet{2018A&A...617A.108A}. We analyse the GOLF time series for the two detectors PM1 and PM2 independently, as well as their mean. Start and end dates of all the time series we used in this work as well as the source of the data are given in Table~\ref{tab:1}.

To study the temporal variation of pseudomode frequencies through the solar cycle, both the GOLF and VIRGO data were split into segments of $100\,\rm d$ in length. The starting point from one segment to the next is moved by $50\,\rm d$, creating a two-time overlap between segments. 

\cite{2005ApJ...623.1215J} calculated the response functions for VIRGO/SPM and GOLF, which characterise the sensitivity of the instruments as a function of atmospheric height. For VIRGO/SPM, the red channel response function peaks approximately 10\,km above the height in the atmosphere that corresponds to an optical depth of unity for a wavelength of 500\,nm $(H_0)$. Then, the green and blue channels' response functions peak a few 10s of km below this height (with the green response function peaking closer to $H_0$ than the blue response function). The response function of GOLF peaks slightly higher in the atmosphere, between 100 and 200\,km above $H_0$, depending on mode of operation. 

\begin{figure}
    \begin{center}
    \includegraphics[width=\linewidth]{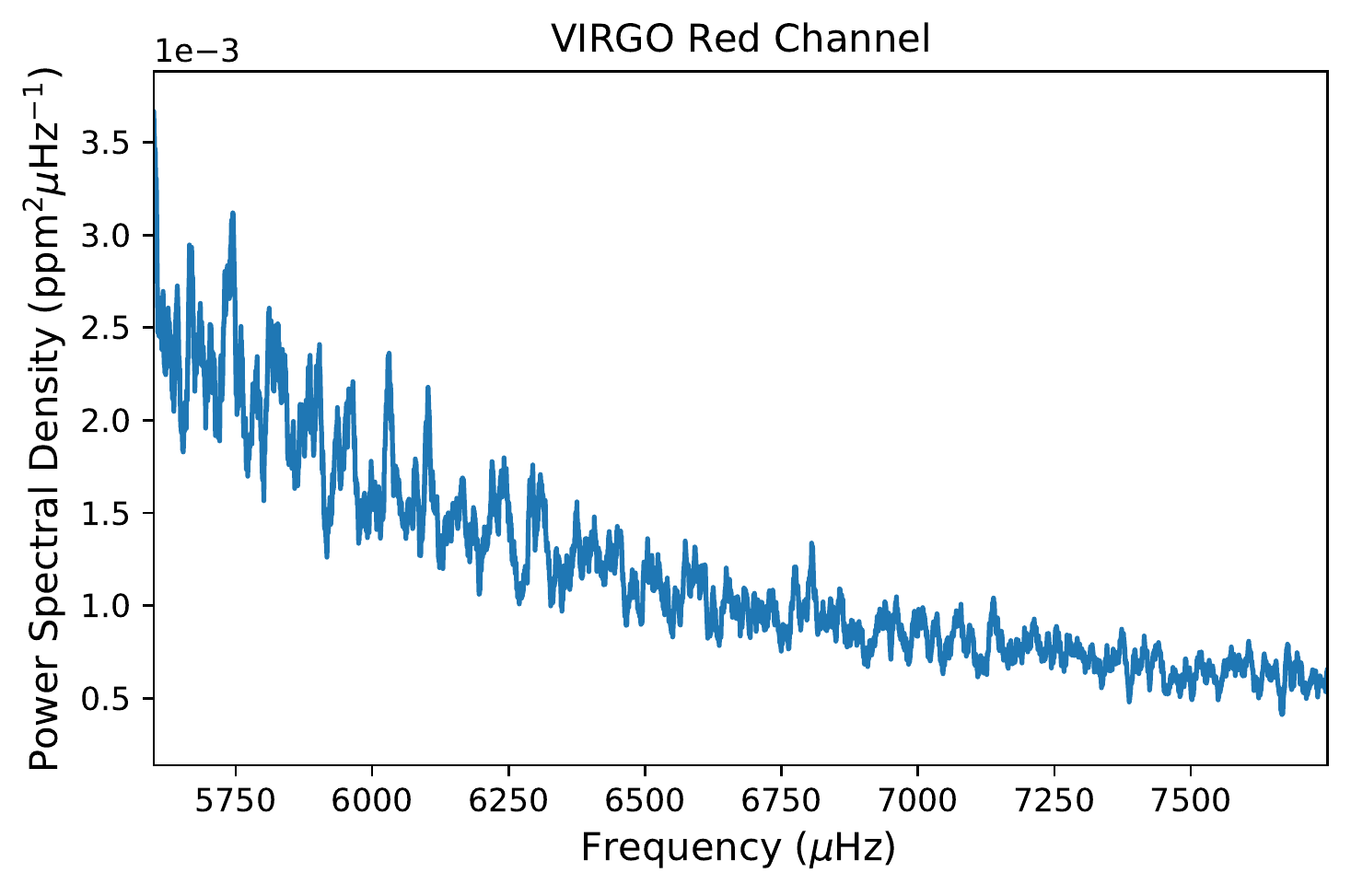}
    \includegraphics[width=\linewidth]{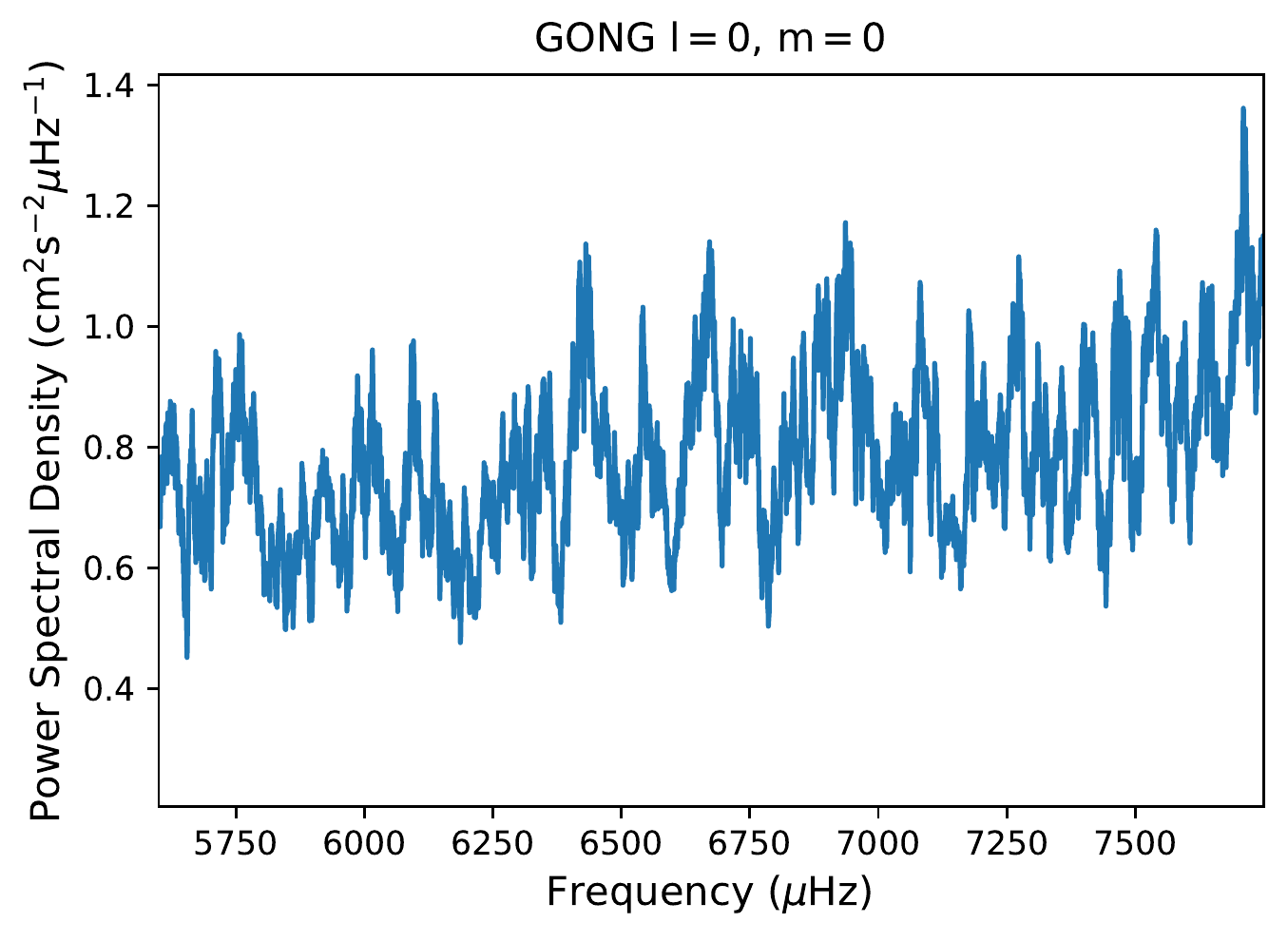}
    \includegraphics[width=\linewidth]{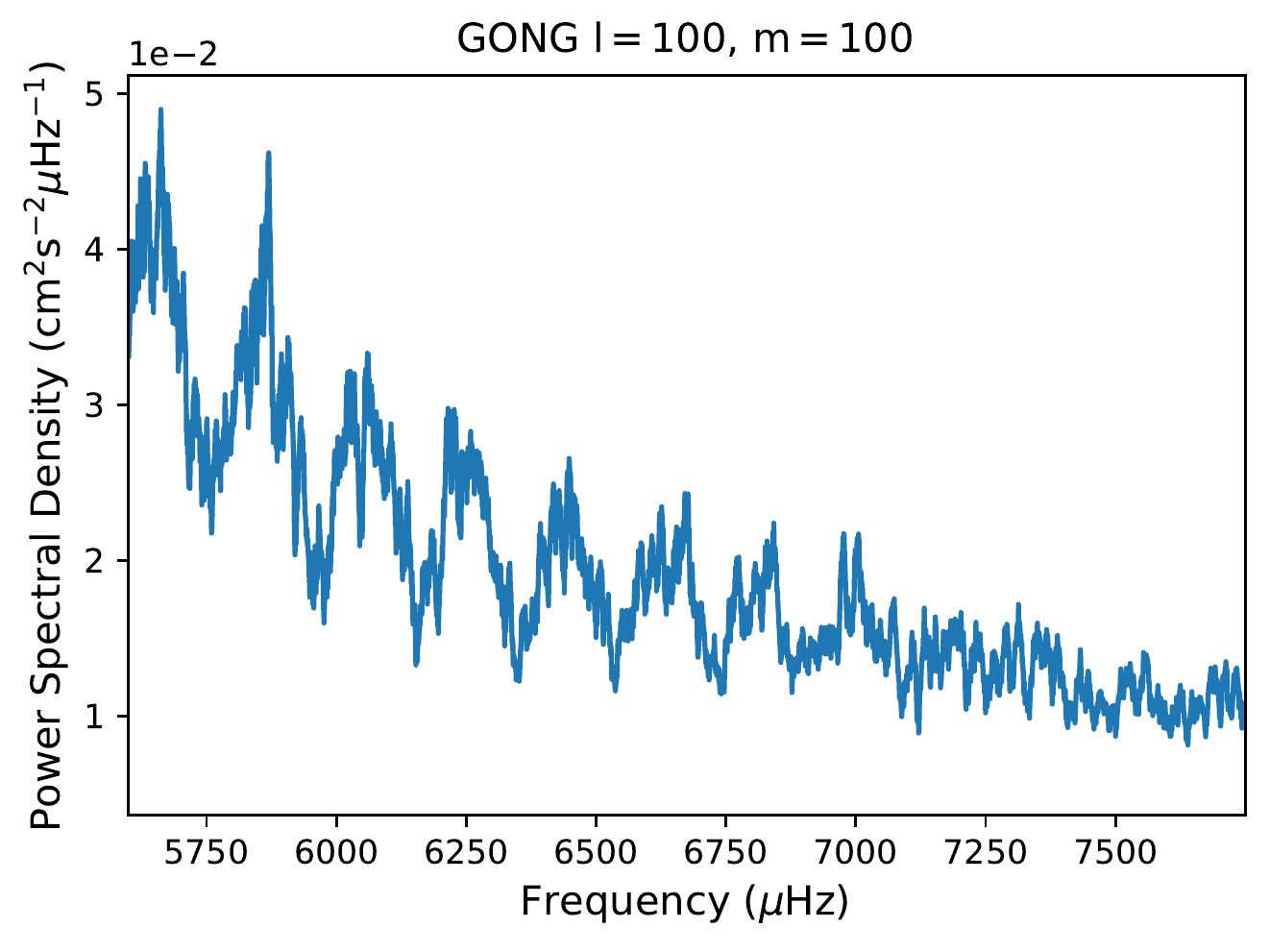}\caption{High-frequency part of the periodograms of the first $200\,\rm d$ of three different time series. Boxcar-smoothing with a width of $10\,\rm\mu Hz$ was applied in all panels. Top panel: VIRGO Red Channel. Middle panel: GONG $l=0$, $m=0$. Bottom panel: GONG $l=100$, $m=0$.} 
    \label{fig:1}
    \end{center}    
\end{figure}

\begin{figure}
    \centering
    \includegraphics[width=\linewidth]{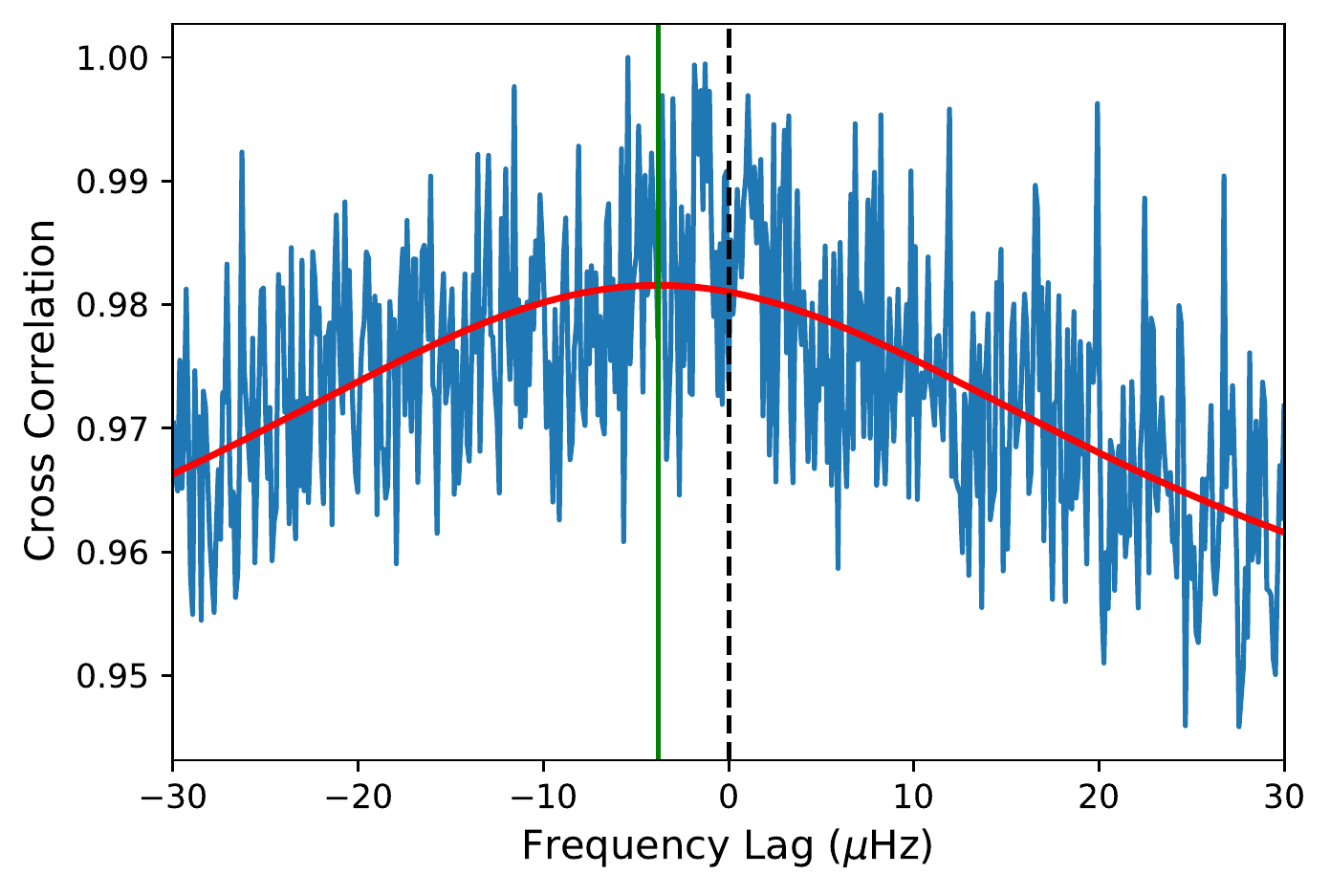}\caption{Cross-correlation function (CCF) of two segments (first segment starting on 16 June 2001, second segment starting on 23 February 2015) from the GONG $l=10$, $m=5$ time series (blue) with a Lorentzian fit (red). The black (dashed) line shows zero lag, while the green (solid) line indicates the maximum of the Lorentzian, i.e., the frequency shift. CCF was normalized to its maximum.}
    \label{fig:2}
\end{figure}

\begin{figure*}
    \includegraphics[width=0.495\linewidth]{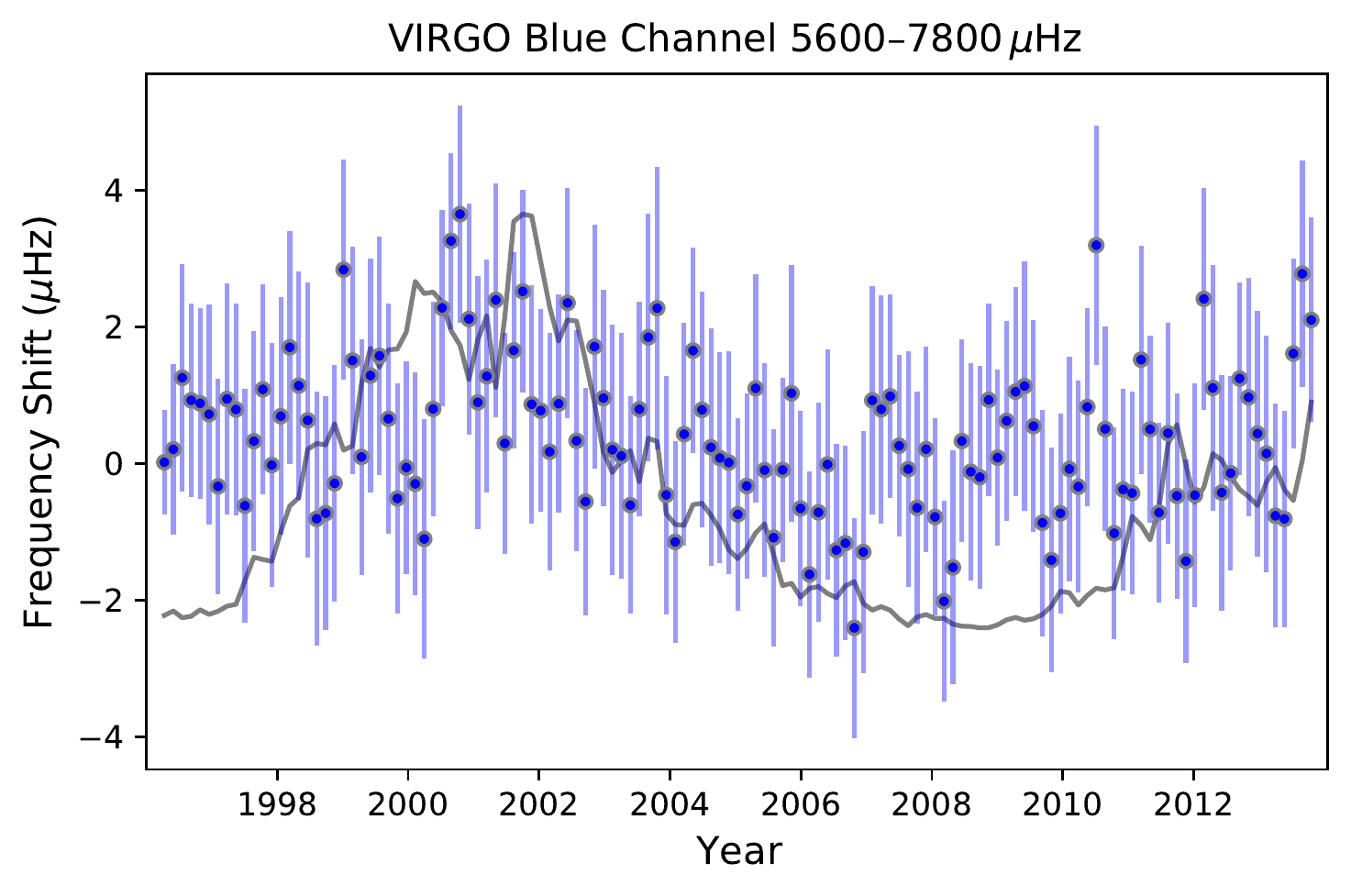}
    \includegraphics[width=0.495\linewidth]{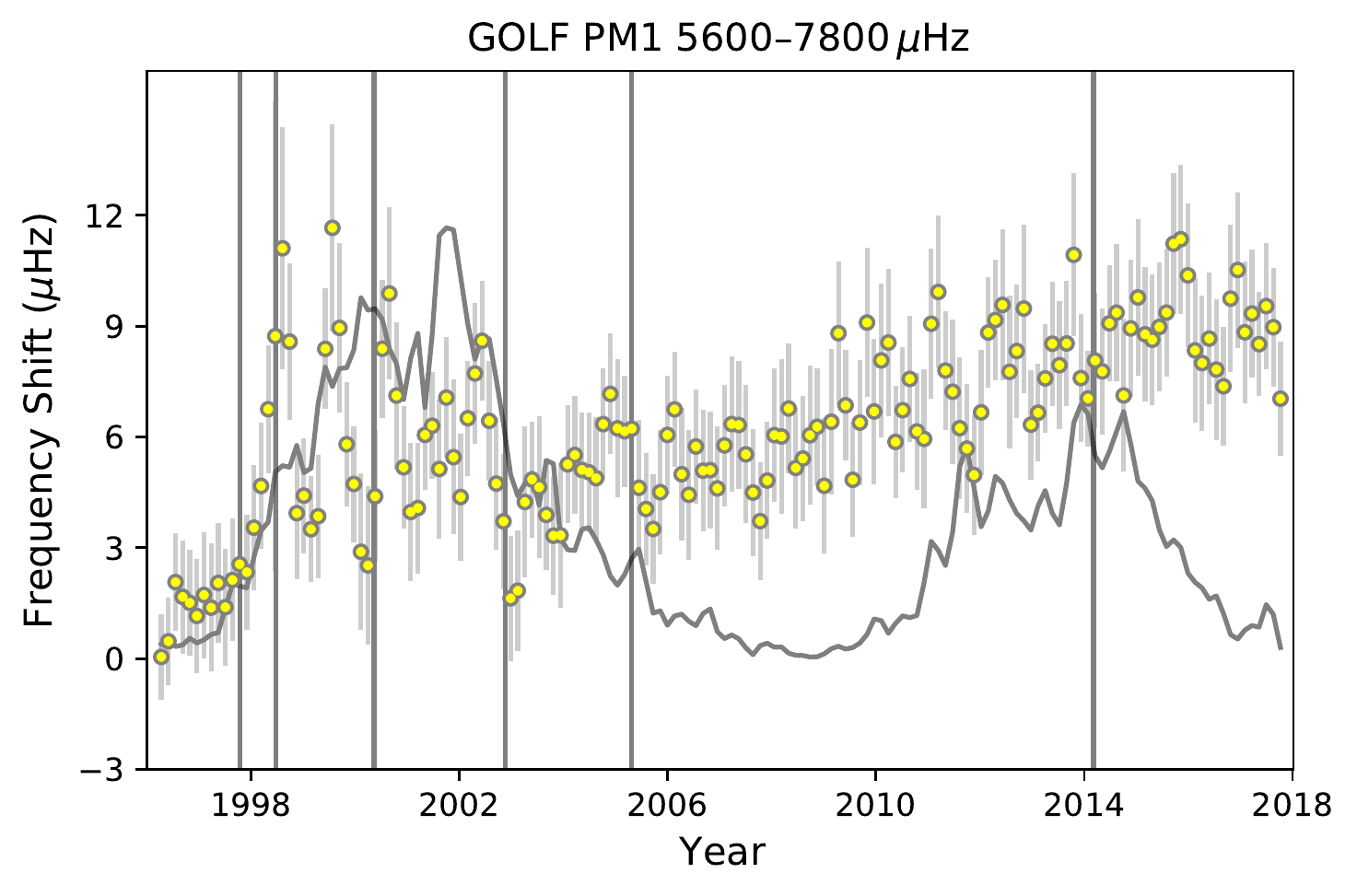}
    \includegraphics[width=0.495\linewidth]{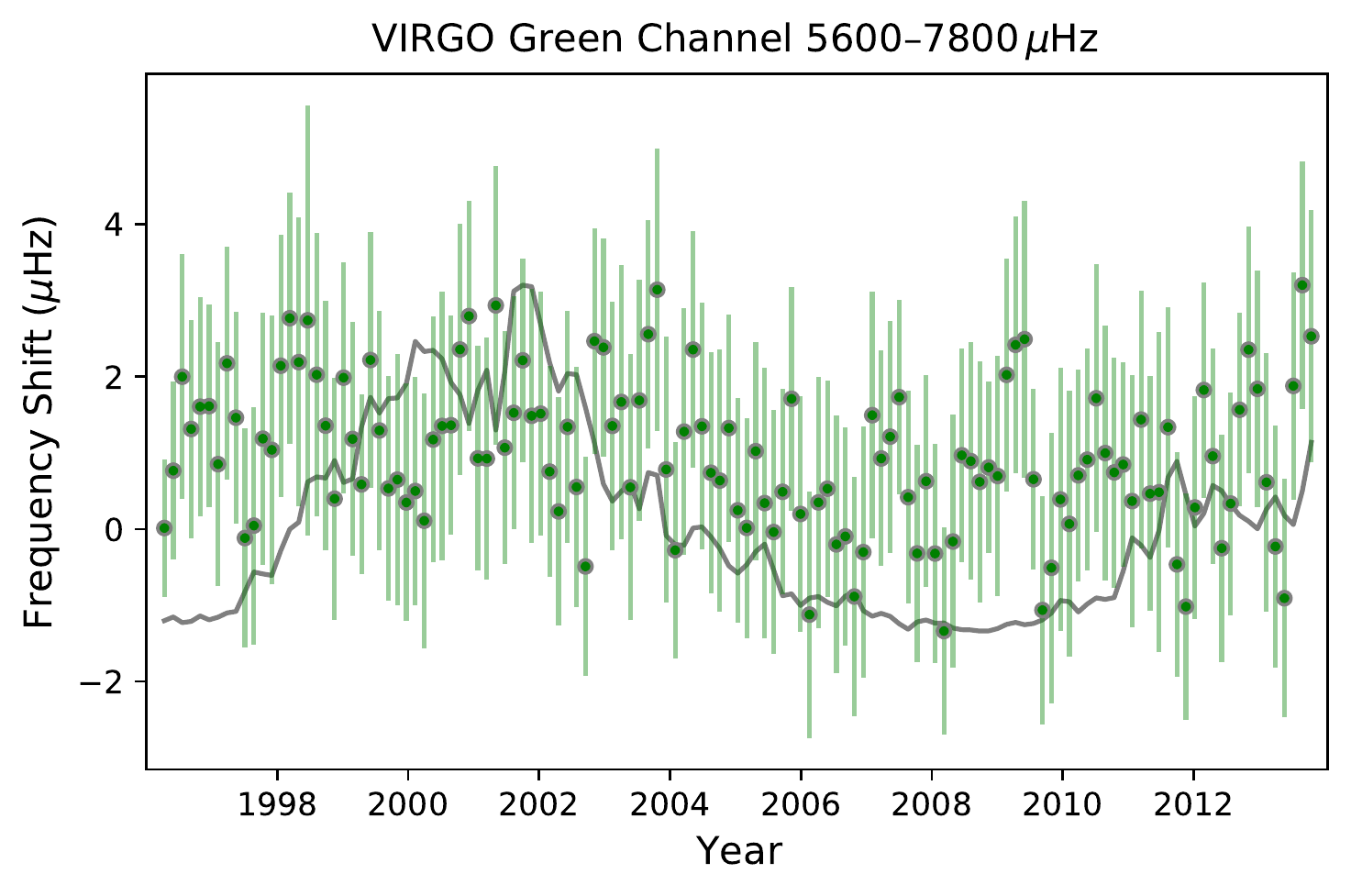}
    \includegraphics[width=0.495\linewidth]{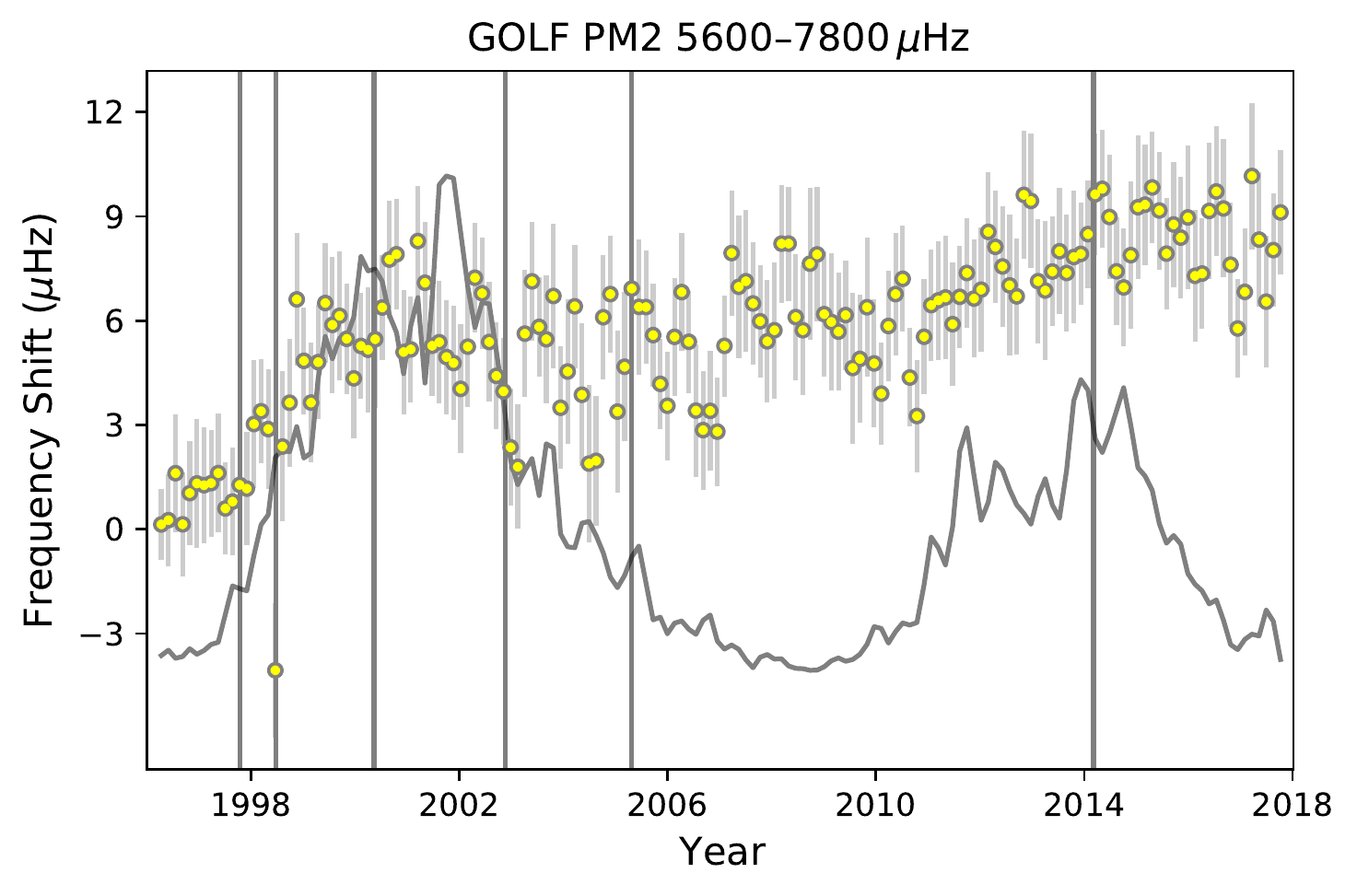}
    \includegraphics[width=0.495\linewidth]{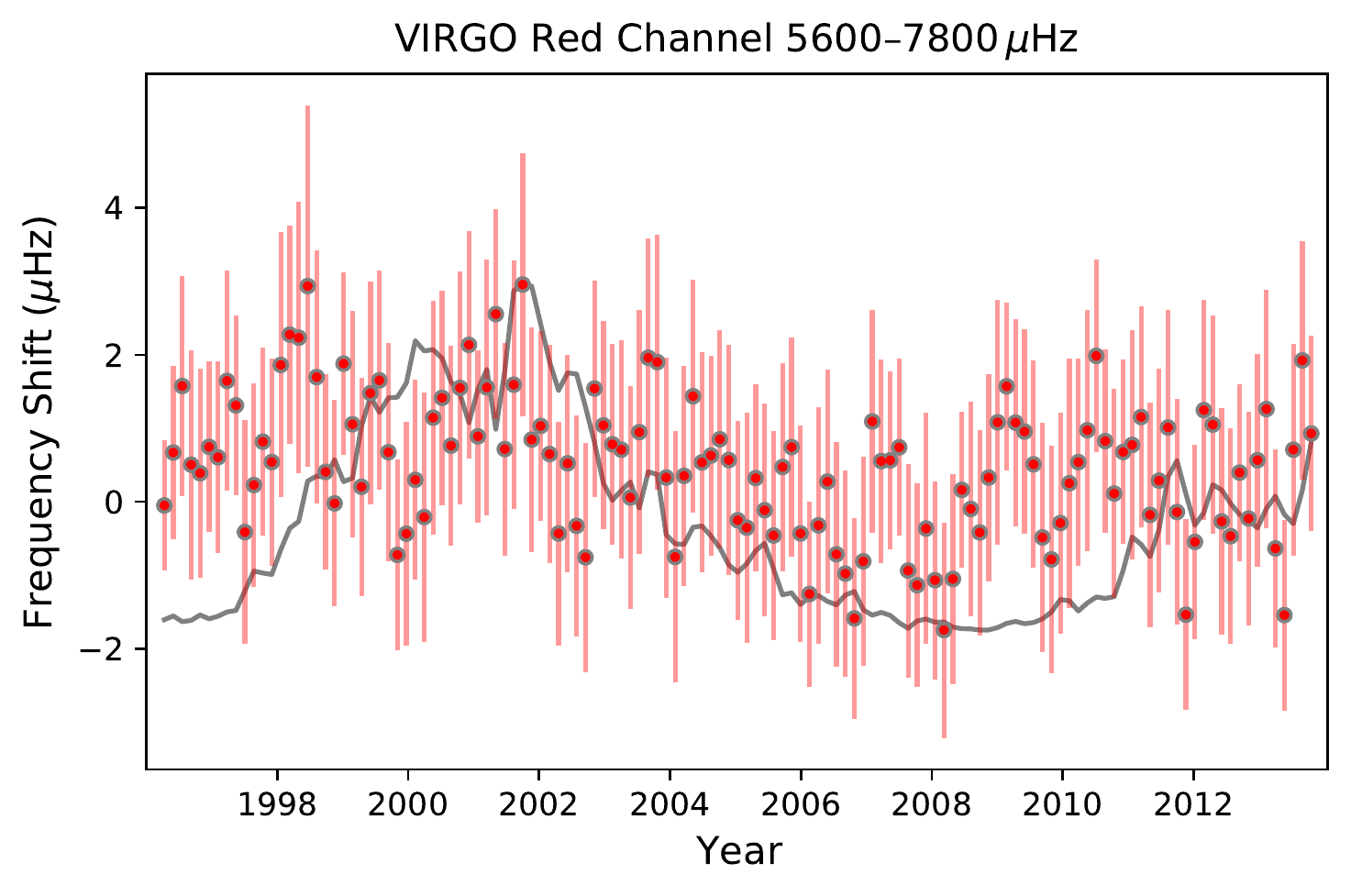}
    \includegraphics[width=0.495\linewidth]{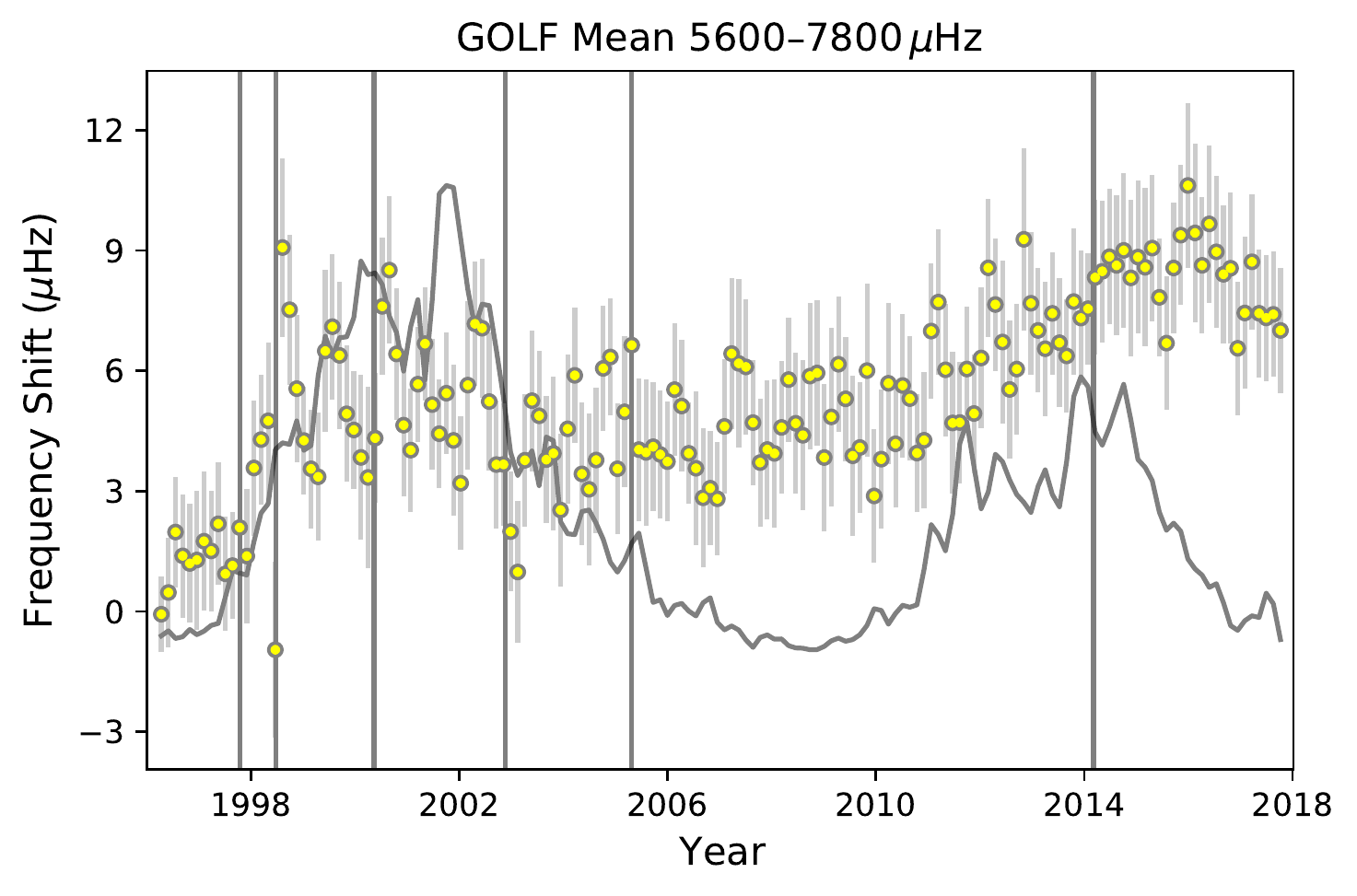}
    \caption{The frequency shifts for VIRGO and GOLF data (coloured data points) with the $F_{10.7}$ index (grey line for all panels) as functions of time. Top left panel: VIRGO Blue Channel frequency shifts (blue data points) with respective $1\sigma$ uncertainties. Top right panel: VIRGO Green Channel frequency shifts (green data points). Bottom left panel: VIRGO Red Channel frequency shifts (red data points). Top right panel: GOLF PM1 frequency shifts. Middle right panel: GOLF PM2 frequency shifts. Bottom right panel: GOLF mean (PM1 \& PM2) frequency shifts. In all right panels the frequency shifts are represented by the yellow data points. }
    \label{fig:3}
\end{figure*}

The Global Oscillation Network Group (GONG) is a network of six sites around Earth that aims to constantly observe the Sun with velocity imagers. We use the gap-filled network-merged 36-day GONG month time series (mrvmt) from the online GONG archive. GONG data can be used to study the oscillations above the acoustic cut-off frequency \citep[e.g.][]{1996Sci...272.1292H}. For GONG, the response function peaks at a height of about $200\,\rm km$ \citep{1989SoPh..120..211J}, which is above GOLF's response function. Spherical harmonic decomposed time series data of the global helioseismology observations from GONG for harmonic degrees $0\leq l\leq 200$ were downloaded covering 16 June 2001 to 15 May 2021. GONG observations before 2001 used $256\times256$ pixel cameras with a cadence of $60\,\rm s$ \citep{1998ESASP.418..209H}. Since the upgrade, the cameras have $1024\times1024$ pixels at the same cadence. Because the differences in resolution are hard to accommodate, we restricted our analysis to start after the network upgrade, i.e. after 16 June 2001. For each azimuthal order available, the 36-day month time series were concatenated to yield time series which covered about 20 years. As with VIRGO and GOLF data, we then separated these long time series into segments of $100\,\rm d$. Again, the starting point from one segment to the next was moved by $50\,\rm d$, creating a two-time overlap between segments.

\begin{table}
	\caption{Start and end dates of the time series used for the analysis.}\label{tab:1}
	\begin{center}
	\begin{tabular}{|c|c|c|c|}
		\hline
		\rule[-1ex]{0pt}{2.5ex} Instrument & Start date & End date & Source \\
		\hline		\hline
		\rule[-1ex]{0pt}{2.5ex} VIRGO & 11 April 1996 & 30 March 2014 & \tablefootnote{\url{http://irfu.cea.fr/dap/Phocea/Vie_des_labos/Ast/ast_visu.php?id_ast=3581}} \\
		\rule[-1ex]{0pt}{2.5ex} GOLF & 11 April 1996 & 10 April 2018 & \tablefootnote{\url{https://www.ias.u-psud.fr/golf/templates/access.html}} \\
		\rule[-1ex]{0pt}{2.5ex} GONG & 16 June 2001 & 15 May 2021 & \tablefootnote{\url{https://nispdata.nso.edu/ftp/TSERIES/vmt/}} \\
		\hline
	\end{tabular}
	\end{center}
\end{table}

To compare any variation in the pseudomode frequencies with the level of solar activity, we used the $F_{10.7}$ index \citep{2013SpWea..11..394T} as a proxy. The $F_{10.7}$ index represents the solar radio flux at a wavelength of $10.7\,\rm cm$. The emission on the solar disk at this wavelength is integrated over one hour to obtain one measurement. The unit of the $F_{10.7}$ index is solar flux units $(\rm sfu)$, with $1\,\rm sfu = 10^{-22}\,\rm Wm^{-2}Hz^{-1}$. \citet{2015SoPh..290.3095B} confirmed that the $F_{10.7}$ index is a good proxy to use when comparing helioseismic data with the level of magnetic activity in the upper chromosphere and the lower corona. Throughout this paper, the $F_{10.7}$ index was rebinned to match the helioseismic data, thus, it was averaged over 100 days with a shift of 50 days to the next segment. To ease comparison, it is included in all figures together with the measured frequency shifts. The $F_{10.7}$ was linearly scaled and shifted to extend from the minimum to the maximum of the frequency shifts presented in each panel. 

\section{Methods}\label{sect:method}
The periodogram was computed for each of the time series segments of the data sets described in Section \ref{sect:data}. We then restricted the periodograms to comprise of just the pseudomode frequency range. We considered two different ranges, namely $5600$--$6800\,\rm\mu Hz$ and $5600$--$7800\,\rm\mu Hz$. The choice of the $5600$--$6800\,\rm\mu Hz$ range was based on the study of \citet{2011JPhCS.271a2029R}, which demonstrated that one can expect modes in this frequency range to be anti-correlated with the solar cycle. The $5600$--$7800\,\rm\mu Hz$ range extends this closer to the Nyquist frequency of $8333\,\rm\mu Hz$ of time series with a cadence of $60\,\rm s$. Figure \ref{fig:1} shows the pseudomode range of periodograms determined from the VIRGO Red channel (top panel), GONG $l=0$, $m=0$ (middle panel) and GONG $l=100$, $m=100$ (bottom panel) time series, smoothed with a boxcar of width $10\,\rm \mu Hz$. In the top and bottom panels, clear modulation of the power spectral density can be observed. These peaks are the pseudomodes we are investigating here. They are less clearly visible in the middle panel. From comparison of the top and bottom panel, it can be appreciated that spacing between consecutive pseudomodes and their individual widths are not constant going from the lowest harmonic degrees to higher harmonic degrees: VIRGO's Sun-as-a-star observations are a superposition of all harmonic degrees where only the lowest harmonic degrees, up to $l\lesssim 3$, are detectable, while the bottom panel is for GONG's $l=100$, $m=100$ time series specifically.

To determine whether the frequencies of the pseudomodes varied with time, we follow the approach of \citet{2017A&A...598A..77K} and \citet{2016A&A...589A.103R}, which was originally used to measure the frequency shifts of p-mode oscillations from solar-like stars using \textit{Kepler} data \citep{2010Sci...327..977B}. This method is based on the cross-correlation between two segments' periodograms of the time series under investigation. The reference segment is kept the same for all calculations and is the following for each instrument: 11 April 1996 -- 20 July 1996 (VIRGO, GOLF), and 16 June 2001 -- 24 September 2001 (GONG). The second segment is then chosen per data set as described in Section \ref{sect:data}. Figure \ref{fig:2} shows such a cross-correlation function (CCF), produced with GONG $l=10$, $m=5$. The CCF is calculated from the reference segment and the segment starting on 23 February 2015. For this figure, the CCF is normalized to its maximum value.

To estimate the frequency shift between two segments and its uncertainty, we employed the resampling approach described in full detail by \cite{2017A&A...598A..77K}. The periodograms of both segments were smoothed using a boxcar smoothing with a width of $10\,\rm\mu Hz$. The first segment of each time series serves as its reference point, as described above. For each segment (reference and segment of interest), a resampled periodogram was generated. For this, a zero-mean normal distribution was multiplied with the square root of the smoothed periodogram. To retain the $\chi_2^2$ distribution of the periodogram, this was done for a real and an imaginary instance, which were then summed and the squared absolute was taken. This yields a new realization of the periodogram which was again $\chi_2^2$-distributed, just as the original. We did this for 100 realizations (for both the reference segment and the segments under investigation) and computed the cross-correlation function (CCF) between these. A Lorentzian function (solid red curve in Figure \ref{fig:2}) was fitted to the CCF of these 100 realizations. For the fit, the CCF was restricted to $\pm 30\,\rm \mu Hz$. This avoided contributions from the first sidelobe in the CCF, yet still well covering the expected range of observable frequency shifts. The Lorentzian function was optimized with a non-linear least-square fit (\textsc{scipy.optimize.curve\_fit}). The centre of the Lorentzian fit was the frequency lag between the two segments' periodograms, which we refer to as the frequency shift. It is indicated by a green solid vertical line in Figure \ref{fig:2}. We calculated the standard deviation of the 100 obtained lags, which was taken as the uncertainty of the frequency shift. The frequency shift between each segment and the reference segment was given by the mean of the 100 realizations in the generated sample.

VIRGO and GOLF take unresolved Sun-as-a-star observations, hence only low harmonic degrees up to $l\lesssim 3$ can be detected in the periodogram. Further, the harmonic degrees cannot be separated as is the case for the resolved observations. In contrast to this, for the GONG data, time series are decomposed by projection of spherical harmonics onto resolved observations for each given $l$ and $m$. Therefore, for the GONG data, the pseudomode frequency shifts as a function of time were determined individually for each $l$ and $m$. We then averaged the frequency shifts over all available $m$ for each harmonic degree. Due to symmetry, spherical harmonic time series contain the same information for $\pm m$, thus only one of them has to be computed from the observations and be considered in the data analysis. This meant that for any given $l$ and any given time, we averaged over a maximum of $l+1$ pseudomode frequency shifts. In the following, we only consider $m$-averaged frequency shifts. This is a big advantage for GONG over VIRGO and GOLF, as for, e.g., $l=100$ a total of $101$ frequency shift results can be averaged. We note that we carry out variance weighted averages when either averaging over azimuthal orders $m$ or subsequently averaging ranges of harmonic degrees.

 We found a correlation between the fill factor of the GONG segments and the measured frequency shifts. The correlation coefficients between the measured shifts and the GONG segments' fill as a function of harmonic degree is included in Fig.~\ref{app:fig:1}. We performed a linear regression of the shifts of every azimuthal order $m$ as a function of fill. This regression utilized the Python module \textsc{statsmodels}'s weighted least squares routine where weighting with the variance of the measured shifts was used. The resulting linear slope was subtracted from the shifts and the extrapolated value at fill$=1$ was added. The uncertainties were properly propagated onto the resulting fill-corrected shifts. In the following, all presented pseudomode frequency shifts have been corrected like this. Figure~\ref{app:fig:2} shows that the linear regression removed most of the correlation between shifts and fill. Being space-based, VIRGO and GONG data typically have $>90\%$ fill. We have thus not accounted for any correlation between fill and pseudomode frequency shift measurements. 

\begin{table}
		\caption{Pearson correlation values $r$, their corresponding $p$ values, Spearman's rank correlation $\rho$, and their corresponding $p$ values, between the instruments' pseudomode frequency shifts and the $F_{10.7}$ index. For GONG, the individual harmonic degrees and averages over harmonic degrees, which are presented in Figures~\ref{fig:4}, \ref{fig:5}, and \ref{fig:7} are listed. Independent, rather than overlapping in time, frequency shift values were used to determine the correlation values.}\label{tab:2}
	\begin{center}
	\begin{tabular}{|c|c|c|c|c|c|}
	\hline
	\rule[-1ex]{0pt}{2.5ex} Instrument & & $r$ & $p$ & $\rho$ & $p$ \\
	\hline\hline
	\rule[-1ex]{0pt}{2.5ex} VIRGO & Blue & 0.34 & 0.006 & 0.31 & 0.02  \\
	\rule[-1ex]{0pt}{2.5ex}  & Green & 0.23 & 0.07 &  0.21 & 0.1  \\
	\rule[-1ex]{0pt}{2.5ex}  & Red & 0.32 & 0.01  & 0.28 & 0.03  \\
	\hline
	\rule[-1ex]{0pt}{2.5ex} GOLF & PM1 & 0.12 & 0.28 & 0.12 & 0.26  \\
	\rule[-1ex]{0pt}{2.5ex}  & PM2 & 0.09 & 0.39  & 0.08 & 0.47 \\
	\rule[-1ex]{0pt}{2.5ex}  & Mean & 0.14 & 0.21  & 0.17 & 0.13  \\
	\hline
	\rule[-1ex]{0pt}{2.5ex} GONG & $0\le l \le 3$ & -0.06 & 0.61 & -0.10 & 0.39 \\
	\rule[-1ex]{0pt}{2.5ex}  & $l=70$ & -0.69 & $<10^{-10}$ & -0.73 & $<10^{-12}$\\
	\rule[-1ex]{0pt}{2.5ex}  & $l=187$ & -0.86 & $<10^{-21}$ & -0.92  & $<10^{-29}$ \\
	\rule[-1ex]{0pt}{2.5ex}  & $0\le l \le 50$ & -0.74 & $<10^{-13}$ & -0.82 & $<10^{-17}$\\
	\rule[-1ex]{0pt}{2.5ex}  & $51\le l \le 100$ & -0.84 & $<10^{-20}$ & -0.90 & $<10^{-26}$\\
	\rule[-1ex]{0pt}{2.5ex}  & $101\le l \le 150$ & -0.87 & $<10^{-22}$ & -0.91 & $<10^{-28}$\\
	\rule[-1ex]{0pt}{2.5ex}  & $151\le l \le 200$ & -0.87 & $<10^{-23}$ & -0.92 & $<10^{-29}$\\
	\hline
    \end{tabular}
	\end{center}
\end{table}

\section{Results}\label{sect:results}
\subsection{VIRGO and GOLF}
Figure \ref{fig:3} shows the measured frequency shifts for all three VIRGO channels (left column from top to bottom: blue, green, and red) and the three GOLF data sets (right column from top to bottom: PM1, PM2, and their mean). In all panels a scaled version of the $F_{10.7}$ index is included (grey curve) as described at the end of Section~\ref{sect:data}. For all data sets included in Figure~\ref{fig:3} we analysed the frequency range $5600$--$7800\,\rm\mu Hz$. We also analysed the more narrow range $5600$--$6800\,\rm\mu Hz$, but the results were only marginally different from the ones presented here. We thus focus on the wider range.

The Pearson correlation coefficients $r$ and the Spearman rank correlation $\rho$ between VIRGO and GOLF frequency shifts and the $F_{10.7}$ index, as well as their corresponding $p$ values, are presented in Table~\ref{tab:2}. There are only marginally significant correlations for the VIRGO channels with the $F_{10.7}$ index. Further, against the expected behaviour \citep[e.g.][]{2011JPhCS.271a2029R}, all correlation values are positive with the strongest correlation found for the VIRGO Blue channel with $r=0.34$ at $p=0.006$. No significant correlation between the measured pseudomode shifts and the $F_{10.7}$ index was observed for any of the three GOLF time series. We note in passing that for the calculation of the correlation values, $r$ and $\rho$, as well as their associated $p$ values, we used independent frequency shift values throughout this paper.

As can be seen from the right column of panels in Figure~\ref{fig:3}, the GOLF pseudomode frequencies appear to drift throughout the time series. We included solid vertical grey lines in these panels, which indicate the times at which either the mode of operation of the GOLF instrument was changed, or the voltage applied to its detectors was increased to counter the effects of ageing. A list of dates of these events is included in the Appendix in Table~\ref{tab:A1}. We propose that the observed pseudomode frequency drift is not of solar origin, but due to these instrumental alterations and possibly also due to the decreasing signal-to-noise ratio as the instrument aged.

\subsection{GONG}
\begin{figure}
    \centering
    \includegraphics[width=\linewidth]{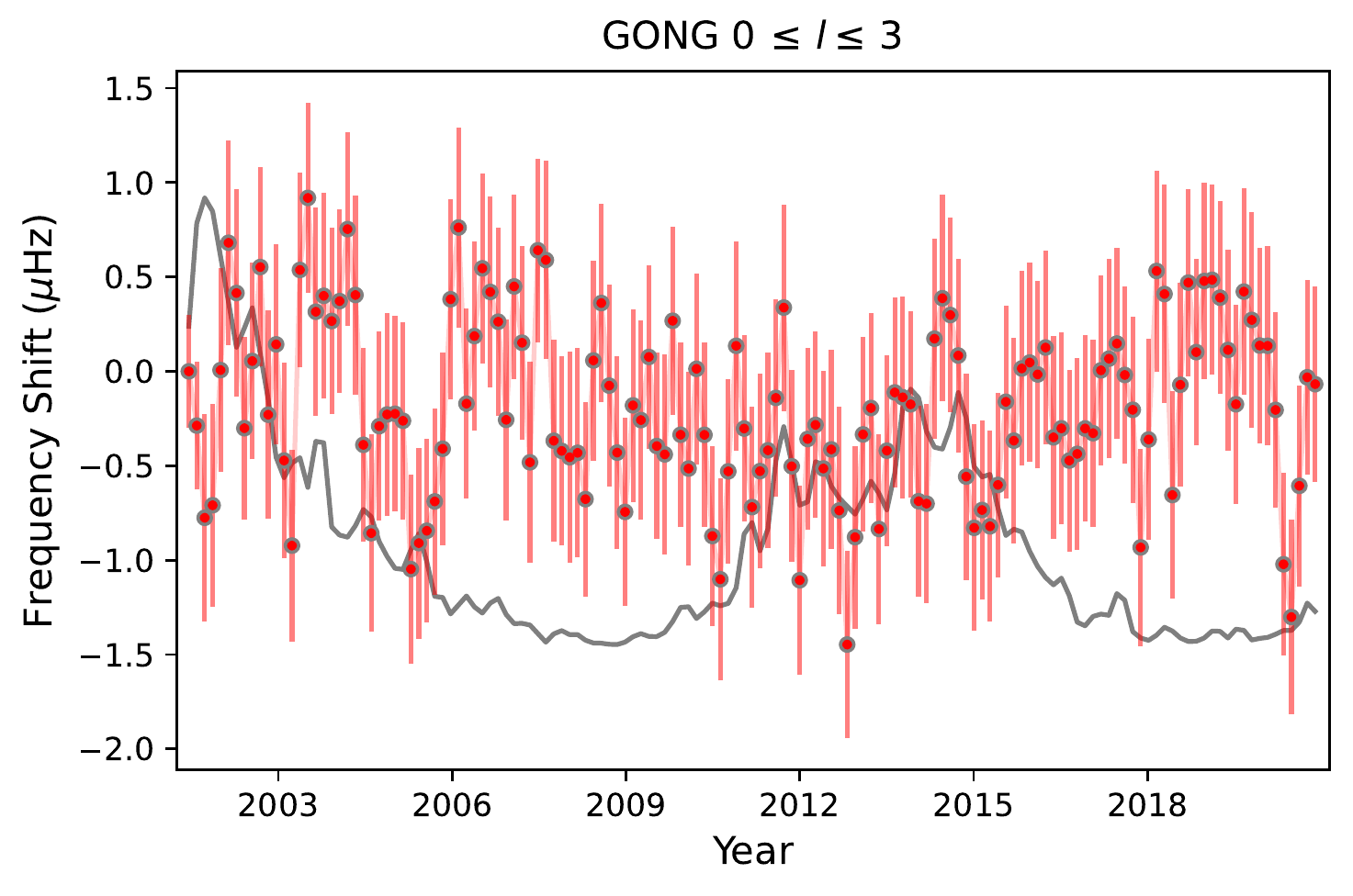}\caption{Frequency shifts as a function of time averaged over harmonic degrees $l=0,1,2$ for the frequency range $5600$--$6800\,\rm\mu Hz$ (red data points). The grey line is a scaled version of the $F_{10.7}$ index for comparison.}
    \label{fig:4}
\end{figure}

To compare GONG data to the VIRGO and GOLF data, we initially consider only the low-degree modes. As shown in Figure \ref{fig:1}, there is little evidence in the $l=0$ spectrum for pseudomodes. For the other low degrees, specifically $l=1, 2$, we also do not see any signs of pseudomodes in the GONG data. Figure \ref{fig:4} shows the average frequency shift for the $0\le l \le 3$ modes. It is again unsurprising that no significant systematic variation in time is observed. The correlation coefficients between the averaged frequency shifts for $0\le l \le 3$ and the $F_{10.7}$ index are $r=-0.06$ with $p=0.61$ and $\rho=-0.1$ with $p=0.39$. Note that the averaging of a mere ten azimuthal orders entering the shifts in Figure~\ref{fig:4} reduced the uncertainties of the frequency shifts compared to those of VIRGO or GOLF, where the different modes are combined into just one time series.

\begin{figure}
    \centering{
    \includegraphics[width=\linewidth]{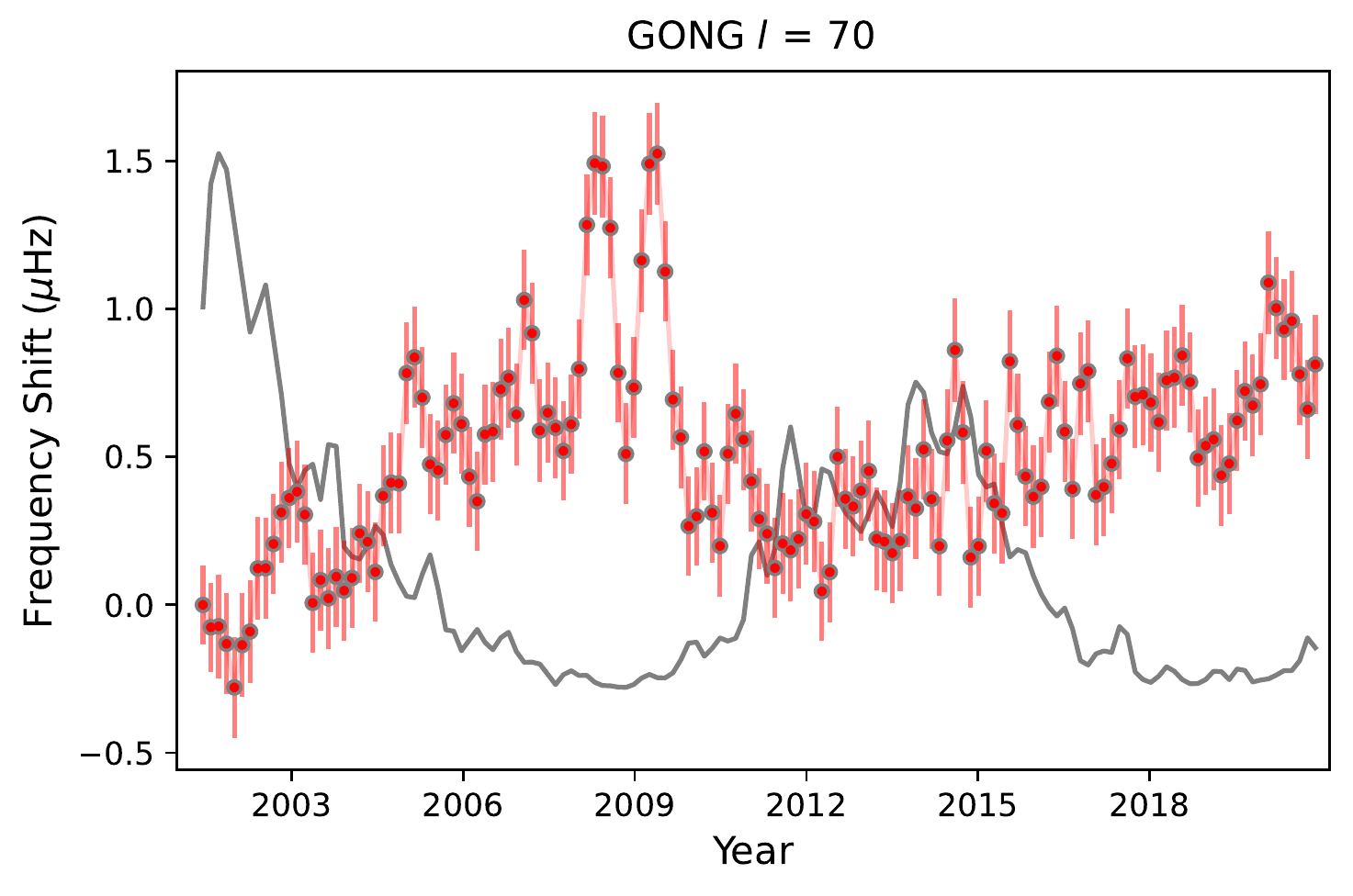}\\
    \includegraphics[width=\linewidth]{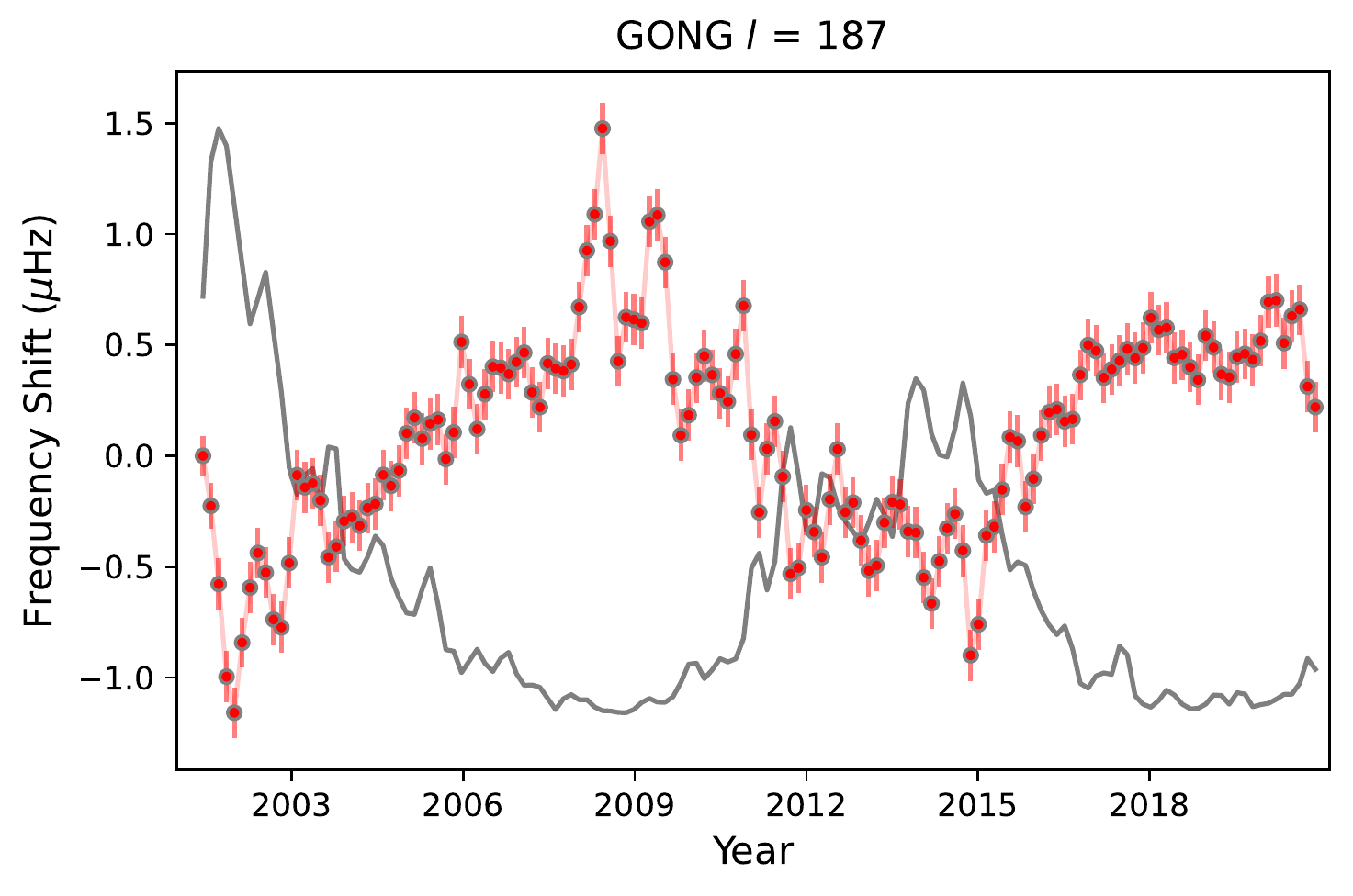}\caption{Frequency shifts obtained from GONG observations as a function of time (red). Frequency shifts were calculated using the frequency range $5600$--$6800\,\rm\mu Hz$. The $F_{10.7}$ index is plotted for comparison (grey). Top panel: Harmonic degree $l=70$. Bottom panel: Harmonic degree $l=187$. }
    \label{fig:5}}
\end{figure}

We now consider the higher degree modes, for which pseudomodes were clearly visible in the periodograms (see bottom panel of Figure \ref{fig:1}). Figure \ref{fig:5} shows the pseudomode frequency shifts observed for two example harmonic degrees, $l=70$ (top panel) and $l=187$ (bottom panel). A scaled version of the $F_{10.7}$ index is included for comparison to the solar cycle (solid grey curve). The behaviour of the pseudomode frequency shifts is similar in both cases. The $l=70$ frequency shifts show slightly larger uncertainties compared to the $l=187$ mode, which is primarily caused by the lower number of azimuthal orders $m$ that enter the weighted average. The correlation coefficients for the $l=70$ mode with the $F_{10.7}$ index are $r=-0.69$ at $p<10^{-10}$ and $\rho=-0.73$ at $p<10^{-12}$. Even though the harmonic degree is only moderately high, the pseudomode frequency shifts are very significant and are clearly in anti-phase with the solar cycle at a very high confidence level. This becomes even more significant for the shifts of the $l=187$ mode, which was chosen here as it is the harmonic degree with the strongest anti-correlation with the $F_{10.7}$ index are $r=-0.86$ at $p<10^{-21}$ and $\rho=-0.92$ at $p<10^{-29}$. 

The amplitude of frequency shifts observed in both the $l=70$ and $l=187$ cases is $\approx 2\,\rm\mu Hz$, which is larger than typically observed for high-frequency p modes, but not inconsistent with an extrapolation of the trend whereby the magnitude of the frequency shift increases with mode frequency \citep[e.g.][]{2017SoPh..292...67B}. This is interesting as p-mode frequency shifts are associated with an internal, near-surface magnetic field but pseudomode frequency shifts are more likely to be associated with an atmospheric magnetic field. \cite{2011JPhCS.271a2029R} have found frequencies of modes above the acoustic cut-off frequency react more strongly to changes in solar activity. We can thus confirm their finding. 

\begin{figure}
    \centering
    \includegraphics[width=\linewidth]{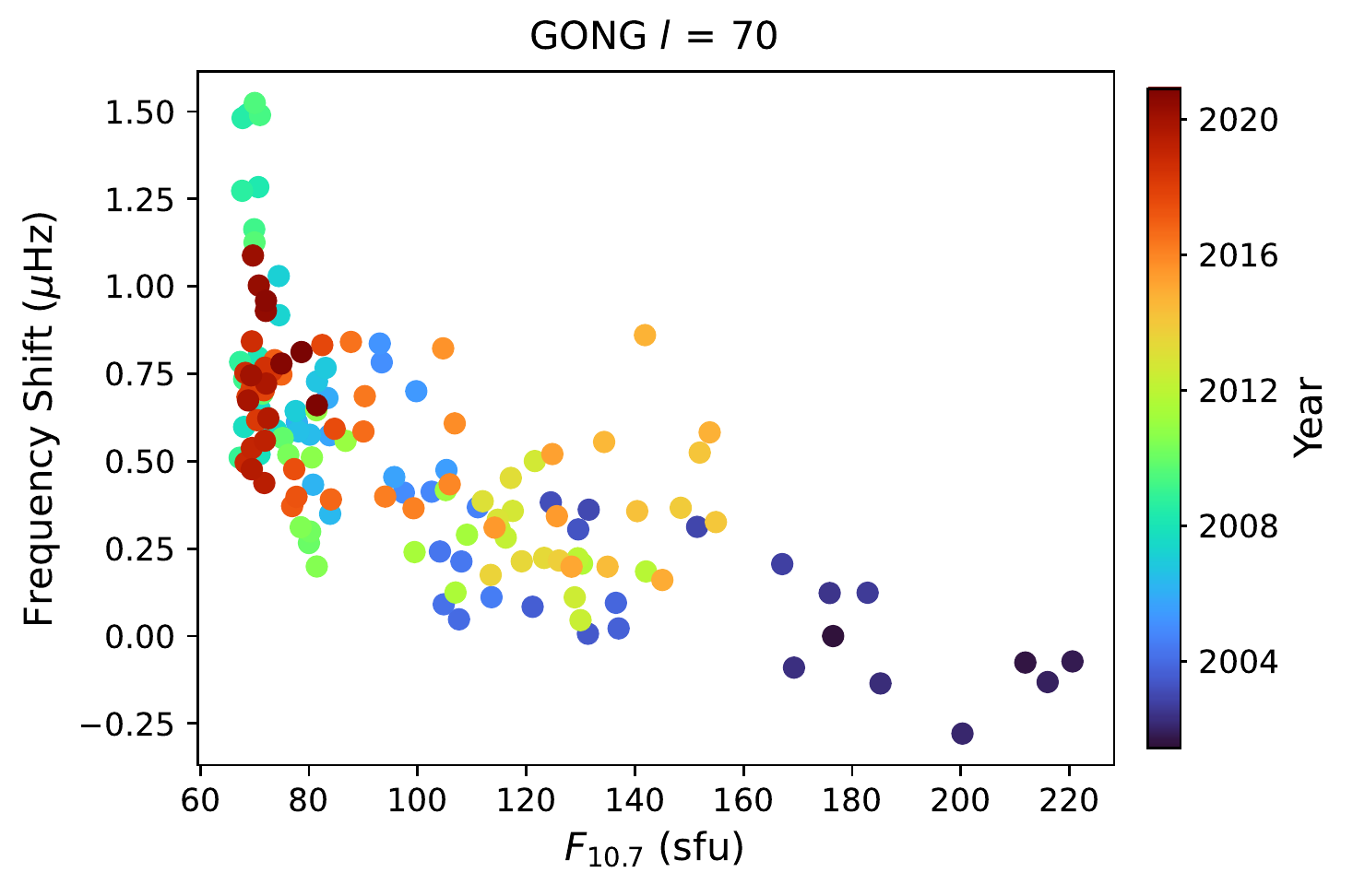}
    \includegraphics[width=\linewidth]{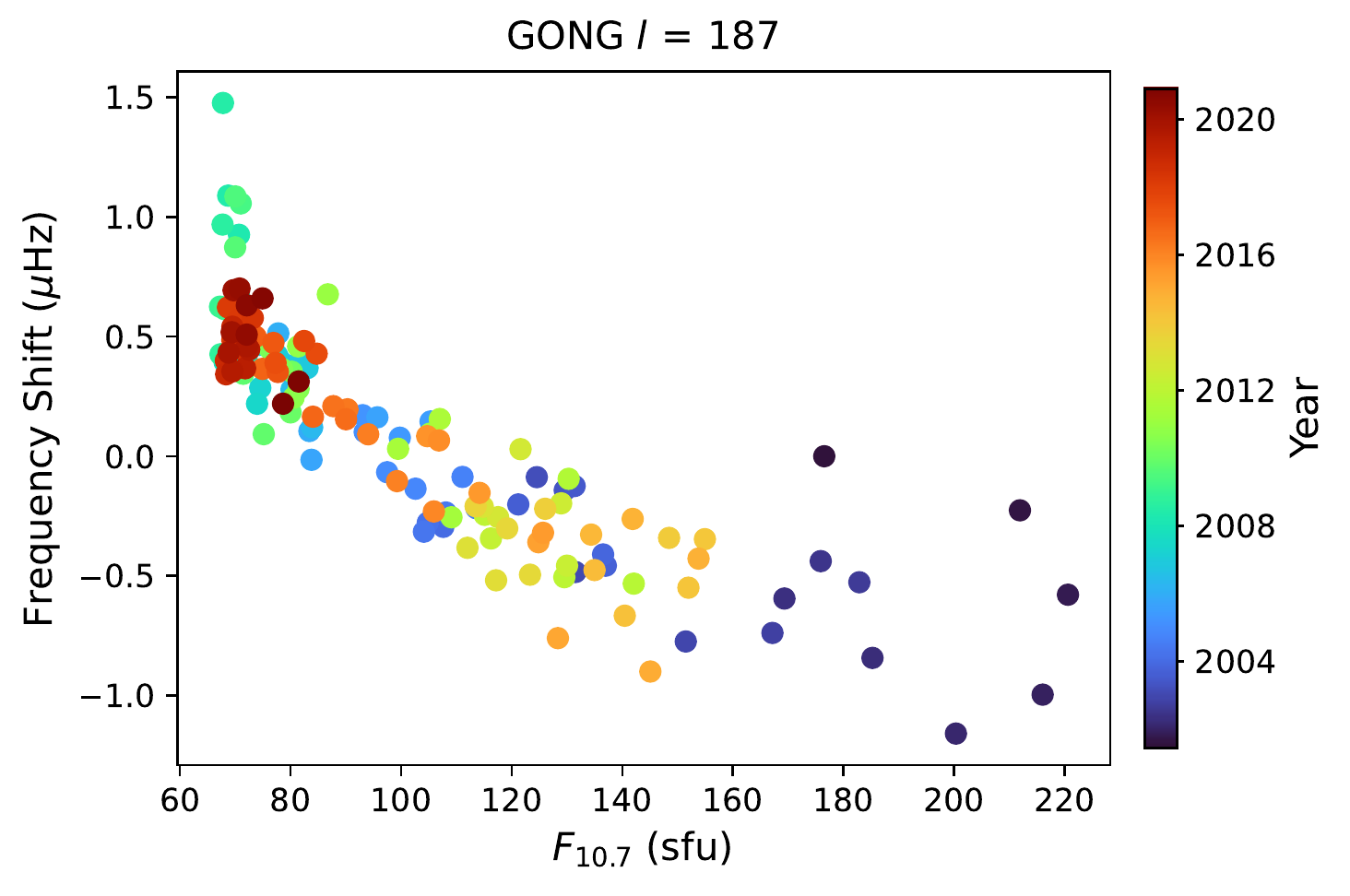}\caption{Frequency shifts for GONG $l=70$ (top panel) and $l=187$ (bottom panel) for the frequency range $5600$--$6800\,\rm\mu Hz$ as a function of the $F_{10.7}$ index.
    }
    \label{fig:8}
\end{figure}

To investigate the anti-correlation between the $F_{10.7}$ index and the pseudomode frequency shifts of these two harmonic degrees further, we plotted one directly against the other. Figure \ref{fig:8} shows this for $l=70$ in the top panel and for $l=187$ in the bottom panel. From these two panels, it can be appreciated that the correlation between the frequency shifts and solar activity is not strictly linear. Thus, for all harmonic degrees and degree ranges considered here and in the following we find $|r|<|\rho|$, see Table~\ref{tab:2}. From the bottom panel of Fig.~\ref{fig:8} it can be seen that the slope of the shifts is steeper below $F_{10.7}\lesssim 80$ than it is for values $80\lesssim F_{10.7}\lesssim 160$. For $l=187$, the slope decreases even further for even higher values of $F_{10.7}$. A similar saturation of mode frequencies with increasing levels of magnetic activity was found for solar p modes by \citet{2019MNRAS.486.1847R}. This may be symptomatic of the fact that acoustic oscillations appear to be more sensitive to small variations in magnetic field through solar minima than other activity proxies, such as $F_{10.7}$ \citep{2017SoPh..292...67B}.
We detect no clear indication of hysteresis in the pseudomode frequencies over the observed time period as has been observed for solar p modes of low harmonic degree \citep{1998A&A...329.1119J} and intermediate degree \citep{2001SoPh..200....3T, 2012A&A...545A..73J}. There is also no evidence of a difference in behaviour between the two different cycles, which has been observed for p modes \citep{2018MNRAS.480L..79H}.

\begin{figure}
    \centering
    \includegraphics[width=\linewidth]{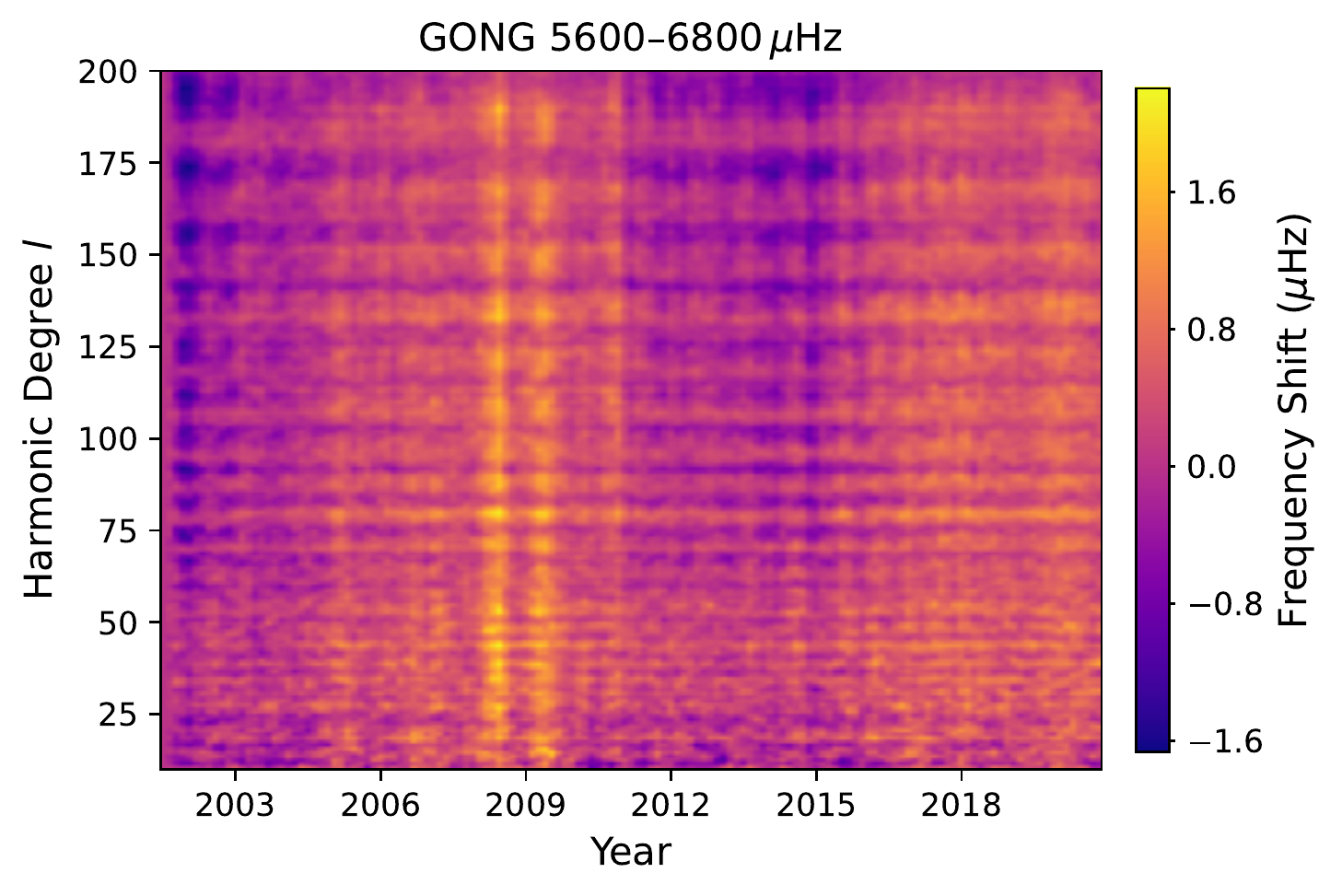}\\
    \includegraphics[width=\linewidth]{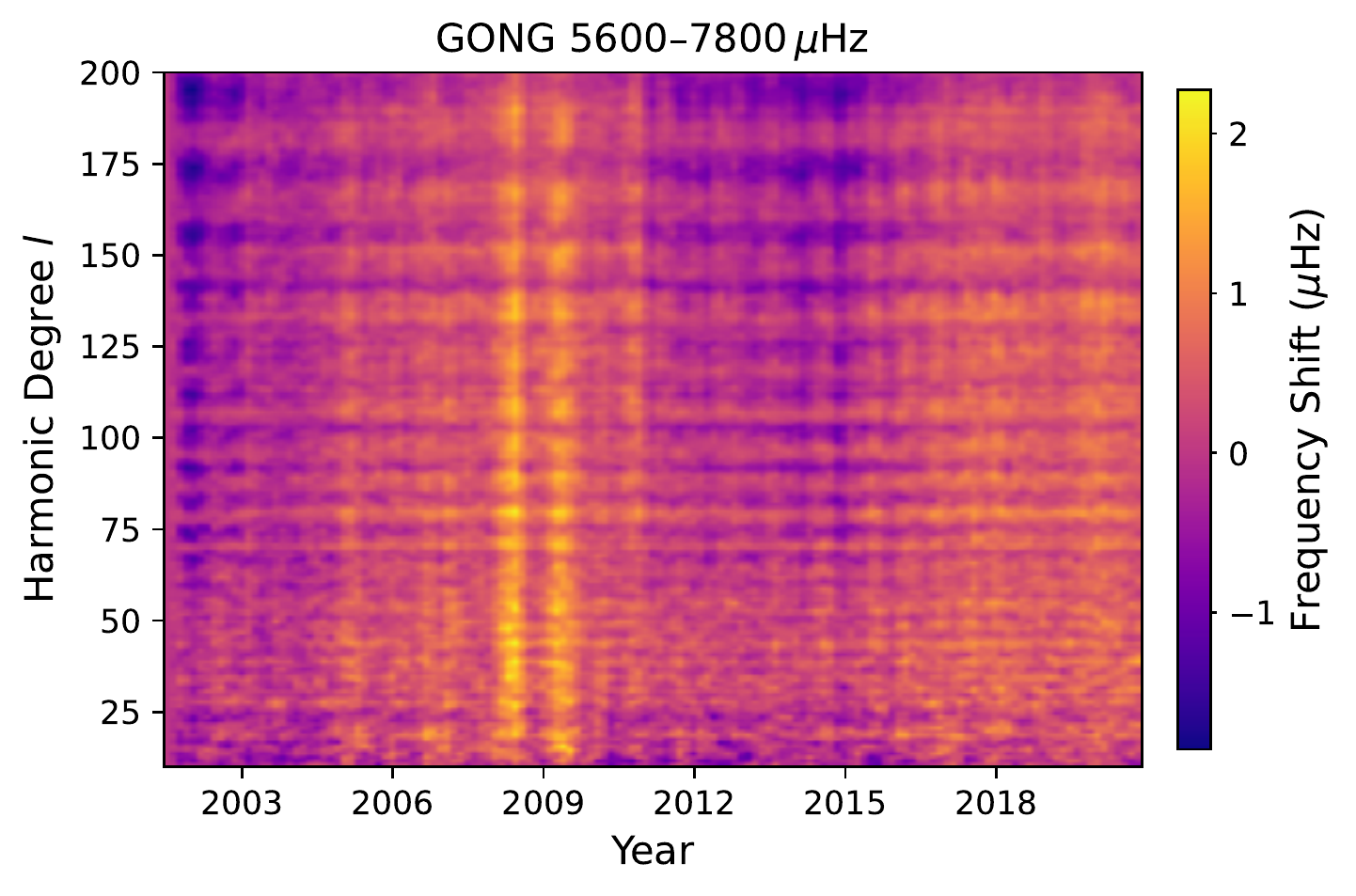}\caption{Frequency shifts as a function of time and harmonic degrees for $10\le l\le 200$. Top: frequency range of $5600$--$6800\,\rm\mu Hz$. Bottom: frequency range of $5600$--$7800\,\rm\mu Hz$.} 
    \label{fig:6}
\end{figure}

Figure \ref{fig:6} displays the pseudomode frequency shifts as a function of time for $10\le l\le 200$. The mean of the lowest harmonic degrees is less well determined due to their smaller number of azimuthal orders. We excluded them here as they would distort the colour bar and impair the presentation of the higher harmonic degrees. The top panel uses the more narrow $5600$--$6800\,\rm\mu Hz$ frequency range and the bottom panel uses the wider range of $5600$--$7800\,\rm\mu Hz$. This way, we can directly compare them and see if we detect any changes in the behaviour of the upper end of the pseudomode frequency range.

In both panels, it is clear that across all $l$ pseudomode frequencies increase starting from the maximum of solar cycle 23 (in 2001) going into the minimum between cycles 23 and 24 (around 2009). At around the time of this minimum, a double peak feature is clearly visible across all $l$ plotted but is not evident for all individual $l$ as can also be appreciated from the two panels of Figure~\ref{fig:5}. This feature coincides with the minimum in the $F_{10.7}$ index, where very little short term variation is observed. It is, however, very reminiscent of the double maximum often seen in solar cycle proxies, which is sometimes referred to as the Gnevyshev Gap and has been associated with the quasi-biennial oscillations \citep[e.g.,][]{2014SSRv..186..359B}. Here, the double peak feature is of unknown origin, but, as far as we can tell, is not a result of any noise or artefact in calculating the frequency shifts. 

There is a clear striation in the shifts through all $l$ that becomes wider at higher $l$. The range of frequency shifts observed appears to be relatively constant over all $l$, with a slight tendency towards larger amplitudes over the cycles for higher $l$. Comparing the two panels, we see that the overall observed pattern of shifts is the same for both frequency ranges. The double peak feature at the extended minimum is clear in both panels. The overall amplitude of the shifts is somewhat larger in the bottom panel. The horizontal striations are also visible for both frequency ranges.

\begin{figure*}
    \centering
    \includegraphics[width=0.495\linewidth]{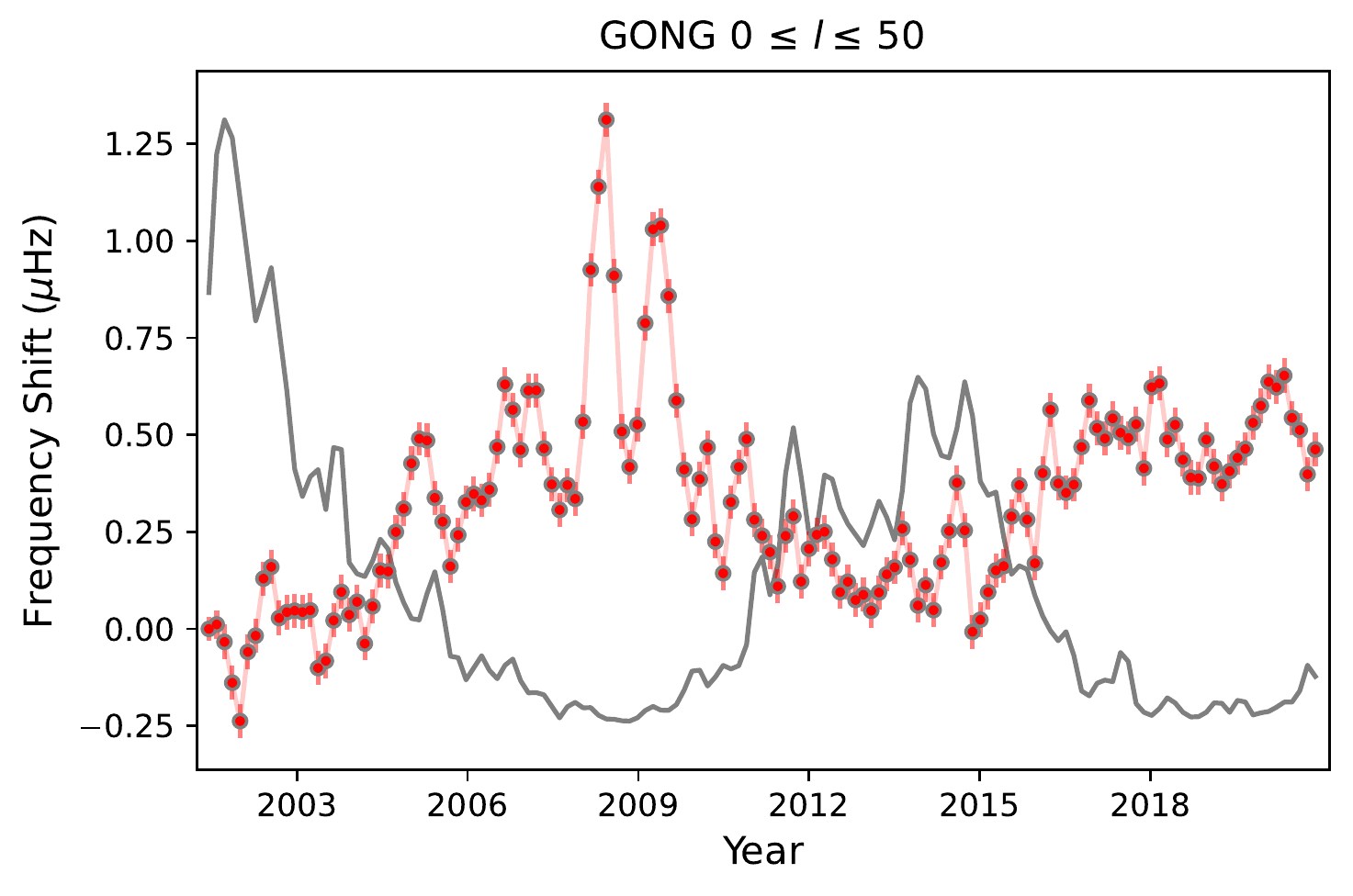}\includegraphics[width=0.495\linewidth]{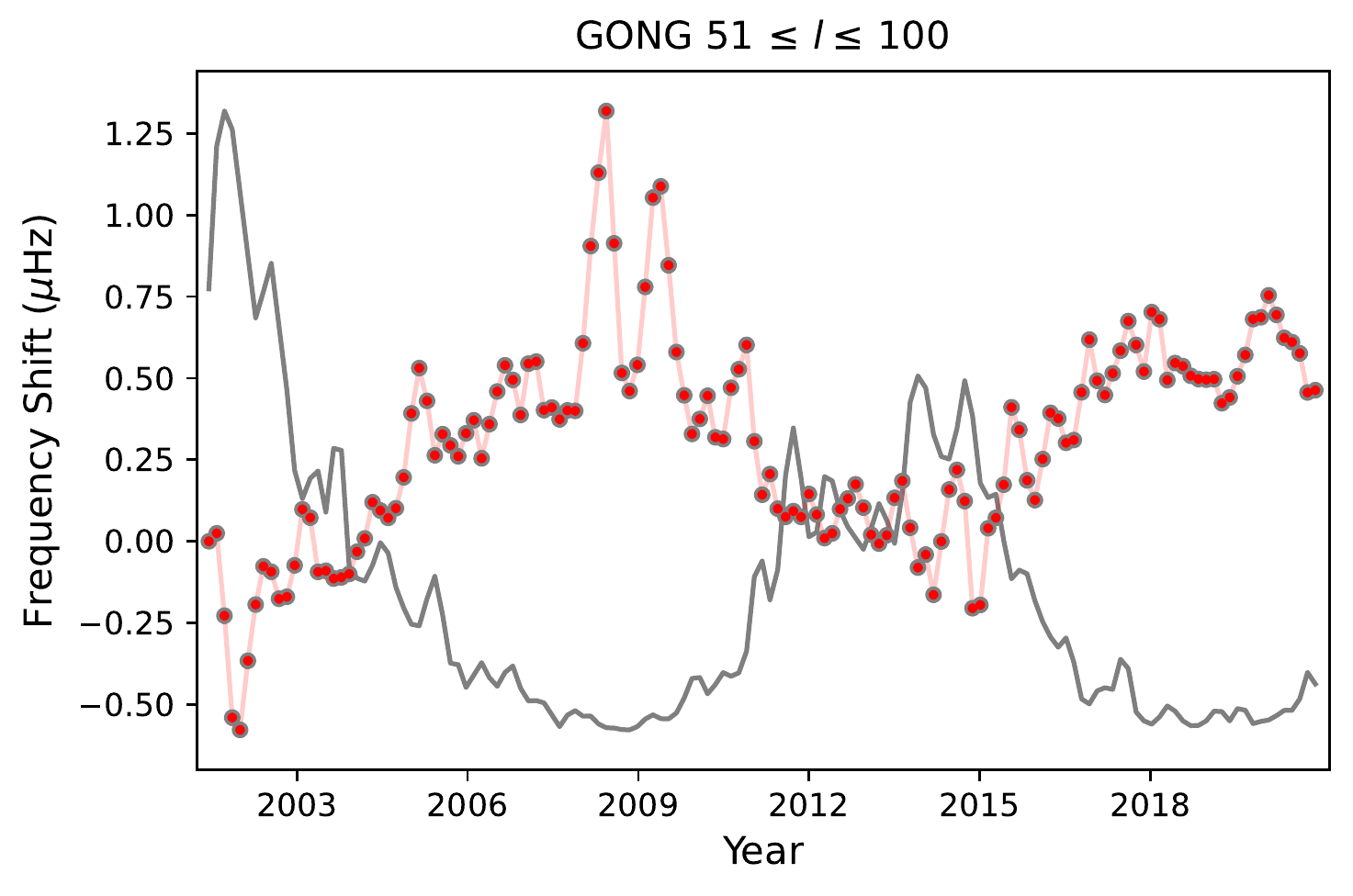}\\
     \includegraphics[width=0.495\linewidth]{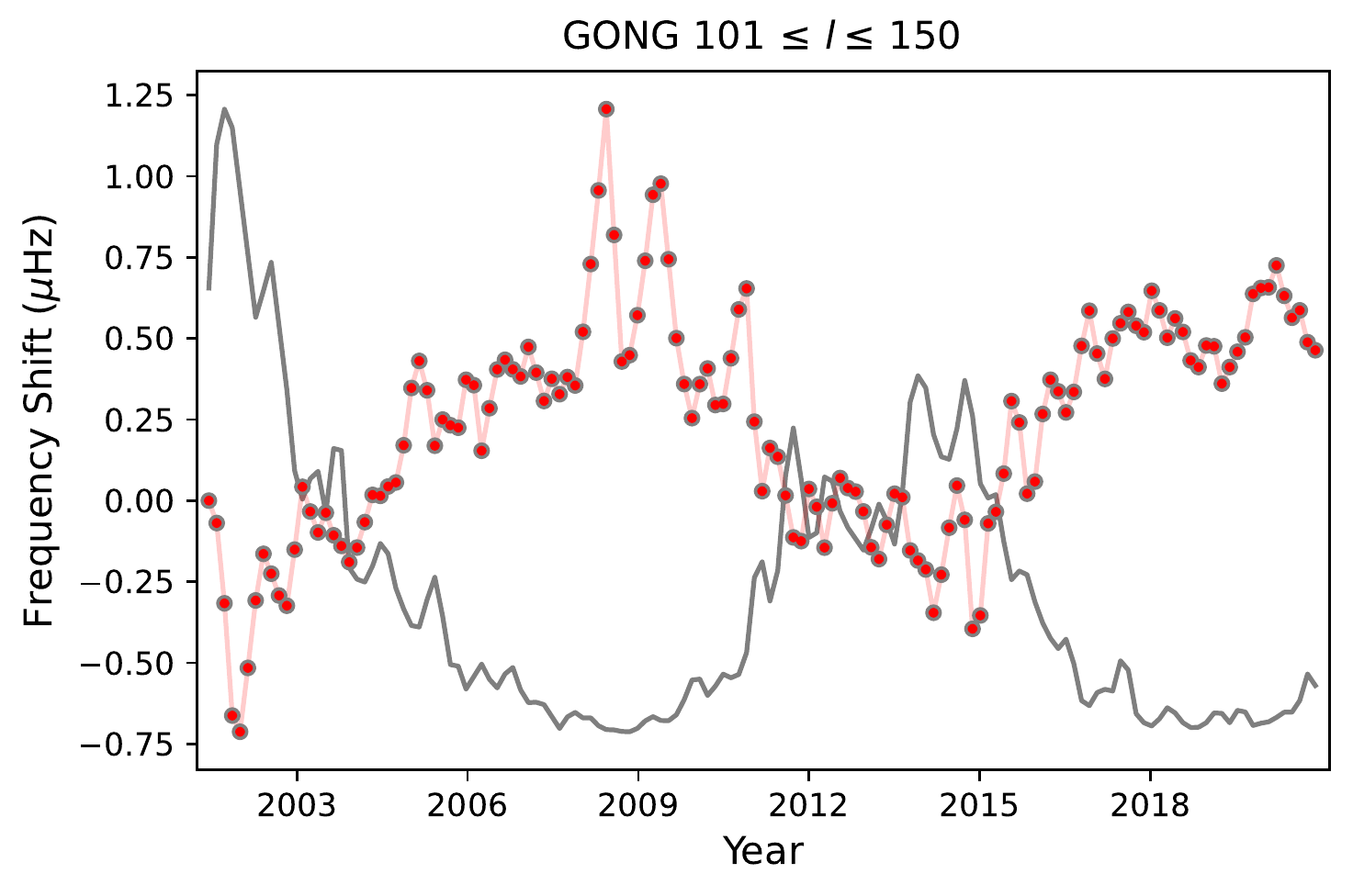}\includegraphics[width=0.495\linewidth]{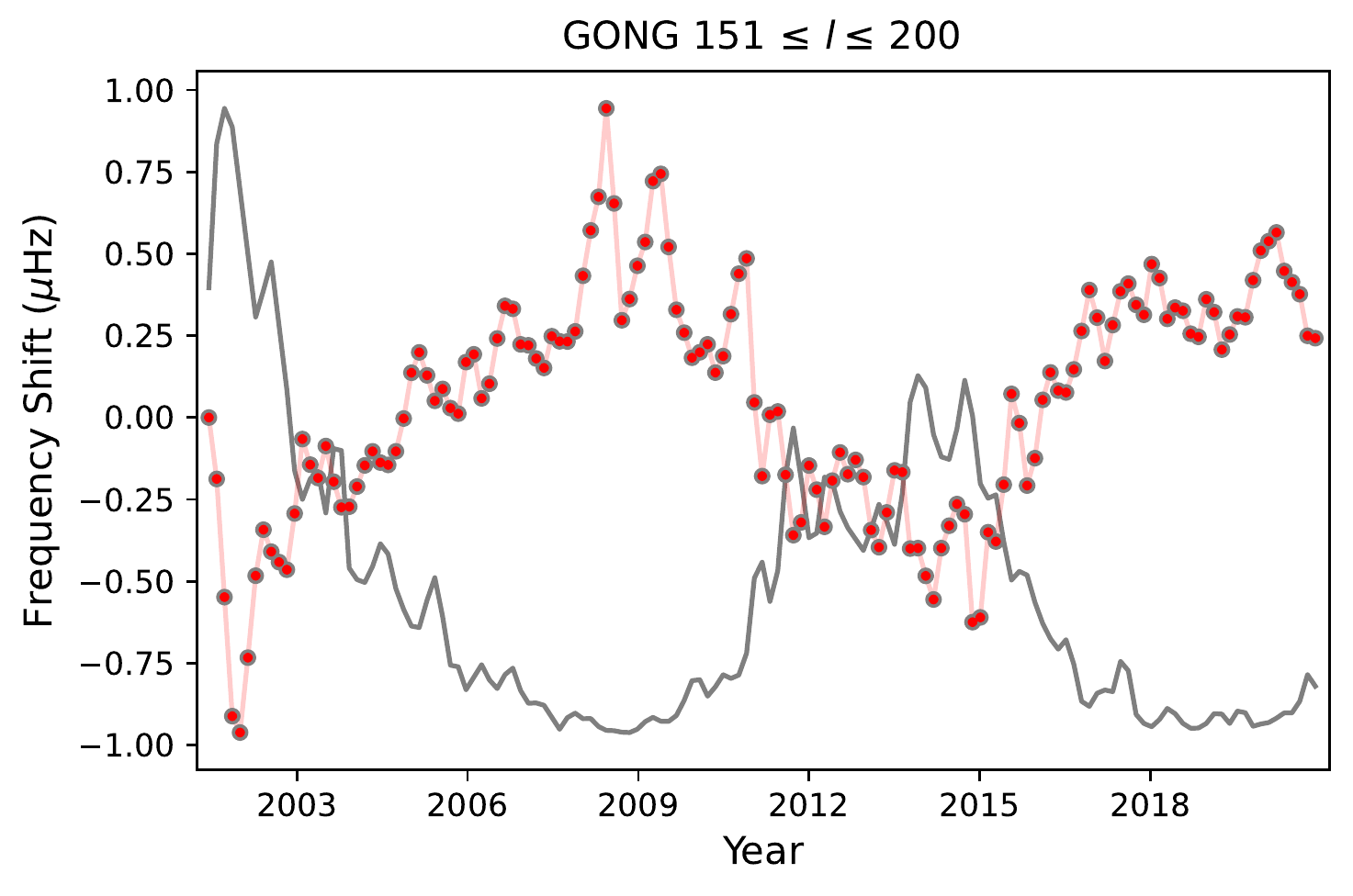}\caption{Frequency shifts as a function of time (red), averaged over different ranges in $l$. The frequency shifts were determined using the frequency range $5600$--$6800\,\rm\mu Hz$. Again the $F_{10.7}$ index is plotted for comparison in grey. Top left panel: $0\le l \le 50$. Top right panel: $51\le l \le 100$. Bottom left panel: $101\le l \le 150$. Bottom right panel: $151\le l \le 200$.}
    \label{fig:7}
\end{figure*}

Since we see similar behaviour for both ranges, all remaining figures for the GONG data will be from the frequency range $5600$--$6800\,\rm\mu Hz$. Similarly, since Figure \ref{fig:6} only varies slowly with $l$, we averaged over different ranges in $l$: $0 \le l \le 50$, $51 \le l \le 100$, $101\le l \le 150$, $151\le l \le 200$. The results are shown in Figure \ref{fig:7}. As expected from Figure \ref{fig:6}, all four panels are visually similar. In all four panels, the pseudomode frequency shifts are clearly anti-correlated with the solar cycle. The double peak feature is most prominent in the low-$l$ data. However, the error bars are marginally smaller for the high-$l$ data (again because of the larger number of $m$ components, which were averaged). The correlation coefficients of all four averages with the $F_{10.7}$ index are listed in Table~\ref{tab:2}. The strongest correlation is found for the $151\le l \le 200$ average with $r=-0.87$ at $p<10^{-23}$ and $\rho=-0.92$ at $p<10^{-29}$.

\begin{figure*}
\centering
  \includegraphics[width=\linewidth]{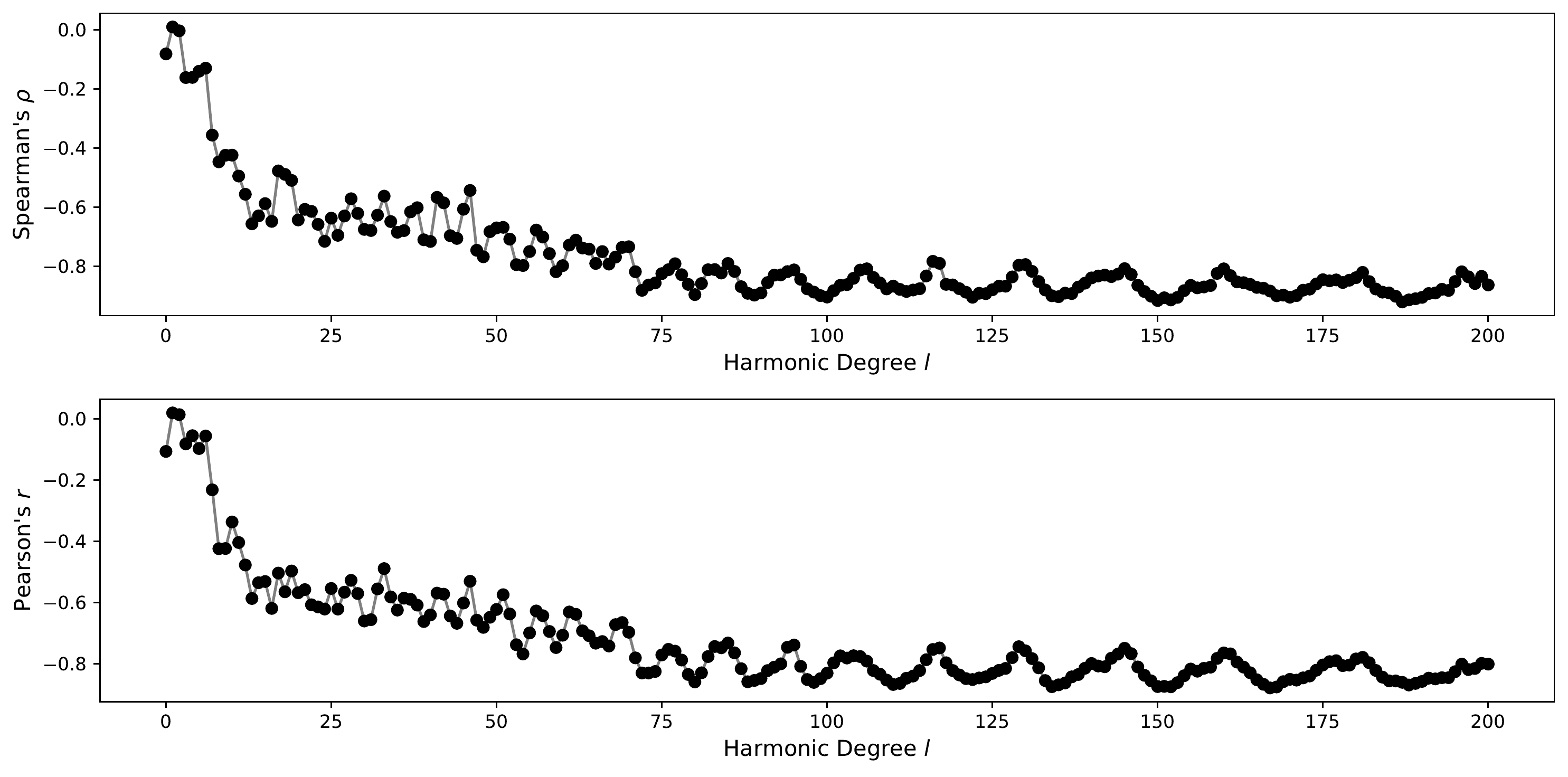}
  \caption{Top panel: Spearman rank correlation $\rho$ between pseudomode frequency shift and $F_{10.7}$ index for GONG data over all $l$. Bottom panel: Pearson correlation coefficient $r$ between pseudomode frequency shift and $F_{10.7}$ index for GONG data over all $l$.}
\label{fig:9}
\end{figure*}

Figure \ref{fig:9} shows the Pearson and Spearman correlation coefficients between pseudomode frequency shifts and the $F_{10.7}$ index over the complete range of harmonic degrees $0\le l\le 200$. The correlation coefficients are negative for all $l>2fa$, highlighting the anti-correlation between the frequency shifts and the solar cycle. As mentioned earlier, the strongest anti-correlation is found for $l=187$. One interesting behaviour of the correlation is that for $25\lesssim l\le 200$ a periodic behaviour is evident. This is most likely related to the striations seen in Figure \ref{fig:6}. There is also the possibility that this feature is related to the leakage matrix or the noise properties of GONG. This could be easily tested by repeating our analysis with data from either SoHO/MDI or SDO/HMI.

\begin{figure*}
    \centering
    \includegraphics[width=0.495\linewidth]{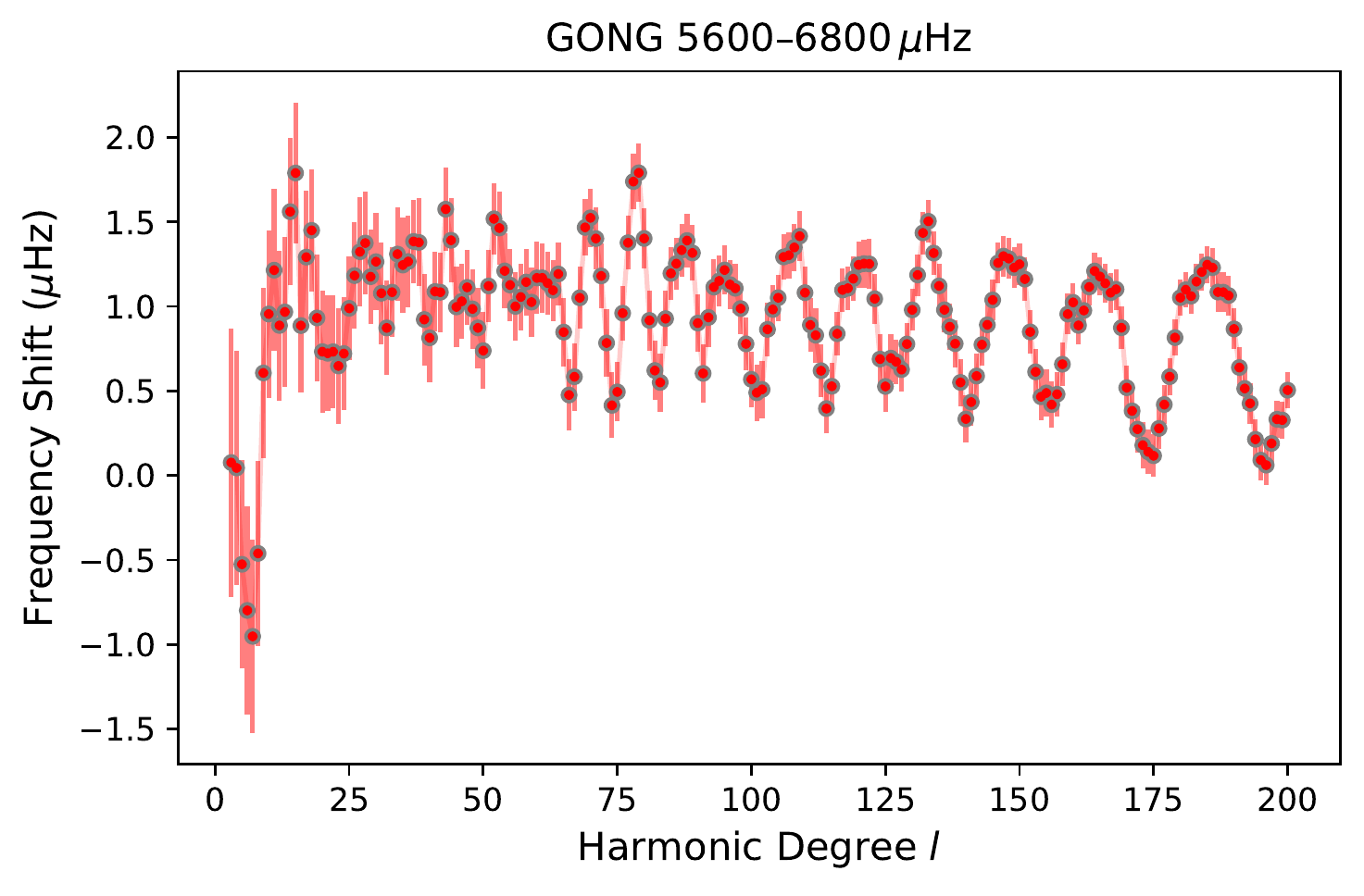}\includegraphics[width=0.495\linewidth]{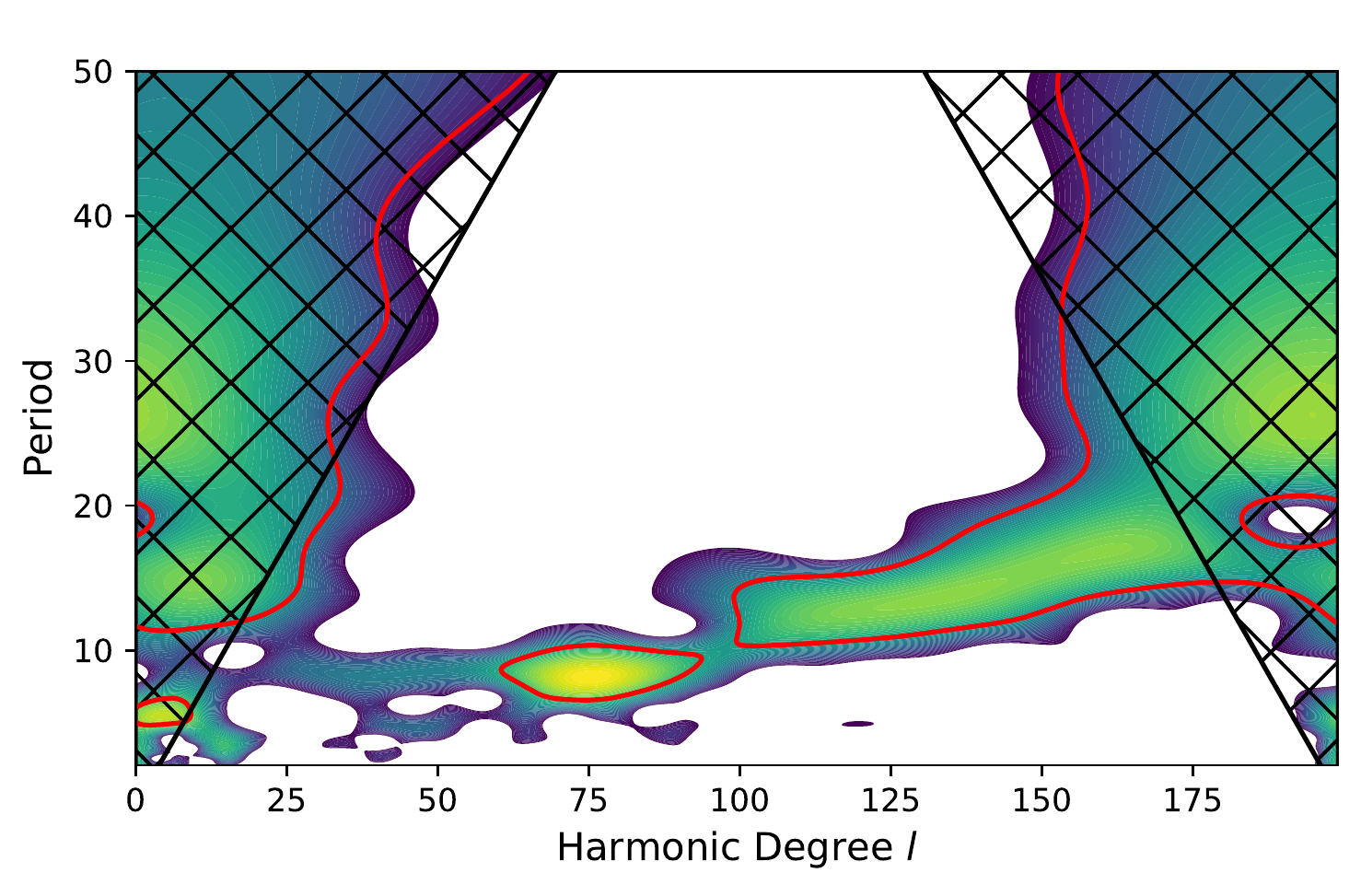}
    \includegraphics[width=0.495\linewidth]{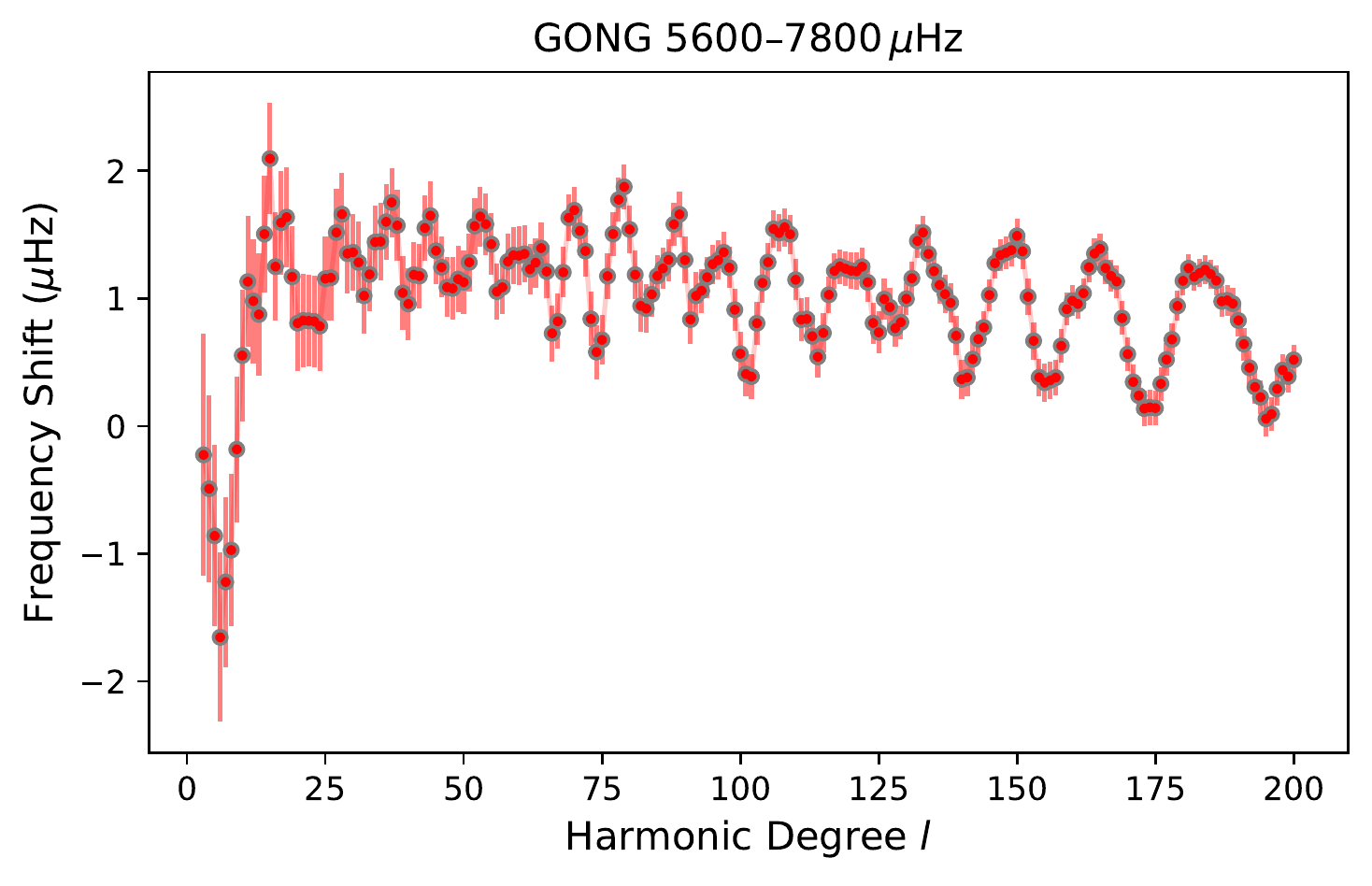}\includegraphics[width=0.495\linewidth]{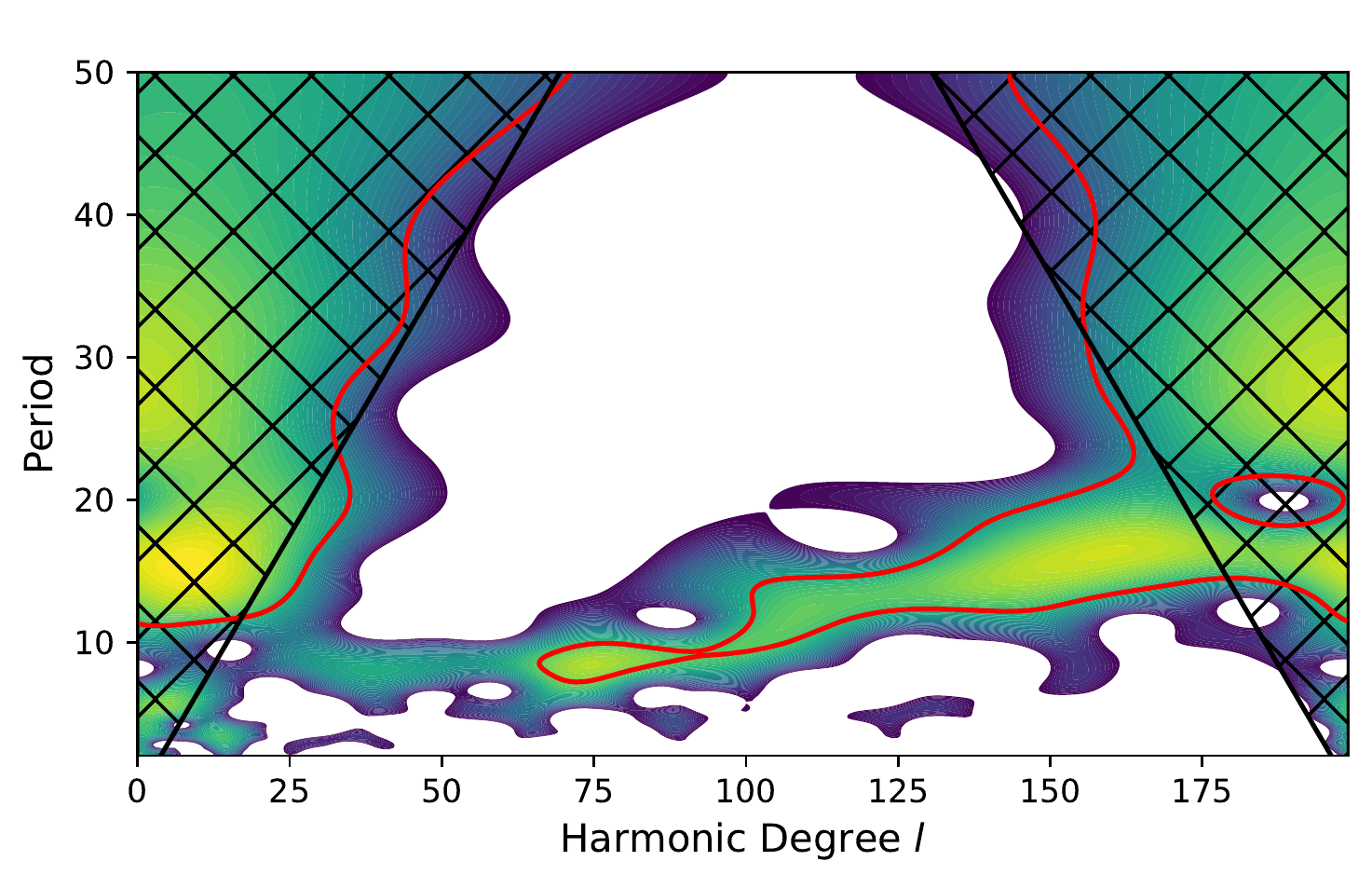}
    \caption{Time slices of the frequency shifts for 28 May 2009 and wavelet analysis. The top panel of each column shows the results for the frequency range $5600$--$6800\,\rm\mu Hz$ and the bottom panel shows $5600$--$7800\,\rm\mu Hz$. Left panels: time slice of the frequency shift as a function of harmonic degree starting at $l=4$. Right panels: wavelet analysis over all harmonic degrees while the period is unitless.} 
    \label{fig:10}
\end{figure*}

To investigate this periodicity and the striations seen in Figure \ref{fig:6} further, we plotted time slices through Figure \ref{fig:6}. The left panels of Figure \ref{fig:10} show one example, namely 28 May 2009, for both frequency ranges. Again, owing to their larger uncertainties and less well-defined shift, we left out the lowest harmonic degrees. In this case, we left out $l<3$. A quasi-periodicity as a function of $l$ is clearly visible in both panels, but the periodicity appears to be non-stationary. Therefore, we also performed a wavelet analysis with a Morlet mother wavelet. The results are shown in the right panels of Figure \ref{fig:10}. The cross-hatching indicates the cone of influence, while the red solid line indicates the 95\% significance level based upon a white noise assumption. The lowest colour level shown is at 80\% significance. A statistically significant period is observed in both frequency ranges, with the significant periodicity extending over a wider range of $l$ in the lower frequency range ($5600$--$6800\,\rm\mu Hz$). It is also clear from Figure \ref{fig:10} that the periodicity increases with $l$, supporting the apparent pattern seen in the left-hand panels of Figure \ref{fig:10}. Although we show only one time period here, this behaviour is typical of all time periods. Such a periodicity has not been observed before and is not predicted by pseudomode frequency shift models \citep[e.g.][]{1996ApJ...456..399J, 1998MNRAS.298..464V}. It is, therefore, imperative that its existence is verified in independent data such as that obtained by Michelson Doppler Imager (MDI), which is also onboard SOHO and the Helioseismic and Magnetic Imager (HMI) on the Solar Dynamics Observatory (SDO).

\section{Summary}\label{sect:conclusions}
We used time series data from VIRGO, GOLF and GONG to measure the frequency shift variations in relation to the solar cycle for solar acoustic oscillations above the acoustic cut-off frequency (pseudomodes). Pseudomodes frequency shifts were not detected at a significant level in the VIRGO, GOLF, or low-degree GONG data. However, they were observed for intermediate-degree GONG data.

We examined two different frequency ranges and found that the behaviour for both ranges of pseudomodes $5600$--$6800\,\rm\mu Hz$ and $5600$--$7800\,\rm\mu Hz$ was similar with the more narrow range giving slightly more significant results. As expected, we found that the frequency shifts are in anti-phase with the solar cycle \citep{2011JPhCS.271a2029R}. We also found a double-peak feature in the pseudomode frequency shifts at solar minimum. Further analysis needs to be done to explain the origin of this feature. There is a periodic behaviour in the pseudomode frequency shifts as a function of $l$. Wavelet analysis of the shifts as a function of harmonic degree at one point in time shows that the period of this quasi-oscillation of the shifts in harmonic degree does increase for larger $l$. This feature was unexpected and such (quasi-)periodic behaviour is not predicted by any of the models for the impact of magnetic fields on pseudomode frequencies. It should be noted that, as we were unable to detect systematic and significant pseudomode frequency shifts in the VIRGO and GOLF data, we are limited to the GONG data set and so the presence of this feature and the double-peak feature should be looked for in other intermediate-$l$ data to verify whether it is solar in origin.

\section*{Acknowledgements}
RK and A-MB acknowledge the support of the Science and Technology Facilities Council (STFC) consolidated grant ST/P000320/1. A-MB acknowledges the support STFC consolidated grant ST/T000252/1.

This work utilizes data from the National Solar Observatory Integrated Synoptic Program, which is operated by the Association of Universities for Research in Astronomy, under a cooperative agreement with the National Science Foundation and with additional financial support from the National Oceanic and Atmospheric Administration, the National Aeronautics and Space Administration, and the United States Air Force. The GONG network of instruments is hosted by the Big Bear Solar Observatory, High Altitude Observatory, Learmonth Solar Observatory, Udaipur Solar Observatory, Instituto de Astrof\'isica de Canarias, and Cerro Tololo Interamerican Observatory. 

The VIRGO and GOLF instruments onboard SoHO are cooperative efforts of scientists, engineers, and technicians, to whom we are indebted. SoHO is a project of international collaboration between ESA and NASA. 

Python wavelet software provided by Evgeniya Predybaylo based on Torrence and Compo (1998) and is available at \url{http://atoc.colorado.edu/research/wavelets/}. This work also uses the Python software libraries \textsc{NumPy} \citep{harris2020array}, \textsc{matplotlib} \citep{2007CSE.....9...90H}, \textsc{pandas} \citep{Pandas2019}, and \textsc{SciPy} \citep{2020SciPy-NMeth}, \textsc{Astropy},\footnote{http://www.astropy.org} a community-developed core Python package for Astronomy \citep{astropy:2013, astropy:2018}, \textsc{statsmodels} \citep{seabold2010statsmodels}, and \textsc{uncertainties} \citep{uncertainties}.

\section*{Data Availability}\label{sec:data_availability}
The data sets were derived from sources in the public domain: The $F_{10.7}$ solar proxy is available online \url{ftp://ftp.seismo.nrcan.gc.ca/spaceweather/solar_flux/}; GONG data can be obtained from \url{https://nispdata.nso.edu/ftp/TSERIES/vmt/}; GOLF data can be found at \url{https://www.ias.u-psud.fr/golf/templates/access.html}; VIRGO data can be downloaded from \url{http://irfu.cea.fr/dap/Phocea/Vie_des_labos/Ast/ast_visu.php?id_ast=3581}. 


\bibliographystyle{mnras}
\bibliography{Astro} 

\begin{thebibliography}{}
\makeatletter
\relax
\def\mn@urlcharsother{\let\do\@makeother \do\$\do\&\do\#\do\^\do\_\do\%\do\~}
\def\mn@doi{\begingroup\mn@urlcharsother \@ifnextchar [ {\mn@doi@}
  {\mn@doi@[]}}
\def\mn@doi@[#1]#2{\def\@tempa{#1}\ifx\@tempa\@empty \href
  {http://dx.doi.org/#2} {doi:#2}\else \href {http://dx.doi.org/#2} {#1}\fi
  \endgroup}
\def\mn@eprint#1#2{\mn@eprint@#1:#2::\@nil}
\def\mn@eprint@arXiv#1{\href {http://arxiv.org/abs/#1} {{\tt arXiv:#1}}}
\def\mn@eprint@dblp#1{\href {http://dblp.uni-trier.de/rec/bibtex/#1.xml}
  {dblp:#1}}
\def\mn@eprint@#1:#2:#3:#4\@nil{\def\@tempa {#1}\def\@tempb {#2}\def\@tempc
  {#3}\ifx \@tempc \@empty \let \@tempc \@tempb \let \@tempb \@tempa \fi \ifx
  \@tempb \@empty \def\@tempb {arXiv}\fi \@ifundefined
  {mn@eprint@\@tempb}{\@tempb:\@tempc}{\expandafter \expandafter \csname
  mn@eprint@\@tempb\endcsname \expandafter{\@tempc}}}

\bibitem[\protect\citeauthoryear{{Appourchaux}, {Boumier}, {Leibacher}  \&
  {Corbard}}{{Appourchaux} et~al.}{2018}]{2018A&A...617A.108A}
{Appourchaux} T.,  {Boumier} P.,  {Leibacher} J.~W.,   {Corbard} T.,  2018,
  \mn@doi [\aap] {10.1051/0004-6361/201833535}, \href
  {https://ui.adsabs.harvard.edu/abs/2018A&A...617A.108A} {617, A108}

\bibitem[\protect\citeauthoryear{{Astropy Collaboration} et~al.,}{{Astropy
  Collaboration} et~al.}{2013}]{astropy:2013}
{Astropy Collaboration} et~al., 2013, \mn@doi [\aap]
  {10.1051/0004-6361/201322068}, \href
  {http://adsabs.harvard.edu/abs/2013A%26A...558A..33A} {558, A33}

\bibitem[\protect\citeauthoryear{{Astropy Collaboration} et~al.,}{{Astropy
  Collaboration} et~al.}{2018}]{astropy:2018}
{Astropy Collaboration} et~al., 2018, \mn@doi [\aj] {10.3847/1538-3881/aabc4f},
  \href {https://ui.adsabs.harvard.edu/abs/2018AJ....156..123A} {156, 123}

\bibitem[\protect\citeauthoryear{{Bazilevskaya}, {Broomhall}, {Elsworth}  \&
  {Nakariakov}}{{Bazilevskaya} et~al.}{2014}]{2014SSRv..186..359B}
{Bazilevskaya} G.,  {Broomhall} A.~M.,  {Elsworth} Y.,   {Nakariakov} V.~M.,
  2014, \mn@doi [\ssr] {10.1007/s11214-014-0068-0}, \href
  {https://ui.adsabs.harvard.edu/abs/2014SSRv..186..359B} {186, 359}

\bibitem[\protect\citeauthoryear{{Borucki} et~al.,}{{Borucki}
  et~al.}{2010}]{2010Sci...327..977B}
{Borucki} W.~J.,  et~al., 2010, \mn@doi [Science] {10.1126/science.1185402},
  \href {https://ui.adsabs.harvard.edu/abs/2010Sci...327..977B} {327, 977}

\bibitem[\protect\citeauthoryear{{Broomhall}}{{Broomhall}}{2017}]{2017SoPh..292...67B}
{Broomhall} A.~M.,  2017, \mn@doi [\solphys] {10.1007/s11207-017-1068-5}, \href
  {https://ui.adsabs.harvard.edu/abs/2017SoPh..292...67B} {292, 67}

\bibitem[\protect\citeauthoryear{{Broomhall} \& {Nakariakov}}{{Broomhall} \&
  {Nakariakov}}{2015}]{2015SoPh..290.3095B}
{Broomhall} A.~M.,  {Nakariakov} V.~M.,  2015, \mn@doi [\solphys]
  {10.1007/s11207-015-0728-6}, \href
  {https://ui.adsabs.harvard.edu/abs/2015SoPh..290.3095B} {290, 3095}

\bibitem[\protect\citeauthoryear{{Broomhall}, {Chatterjee}, {Howe}, {Norton}
  \& {Thompson}}{{Broomhall} et~al.}{2014}]{2014SSRv..186..191B}
{Broomhall} A.~M.,  {Chatterjee} P.,  {Howe} R.,  {Norton} A.~A.,   {Thompson}
  M.~J.,  2014, \mn@doi [\ssr] {10.1007/s11214-014-0101-3}, \href
  {https://ui.adsabs.harvard.edu/abs/2014SSRv..186..191B} {186, 191}

\bibitem[\protect\citeauthoryear{{Chaplin}, {Elsworth}, {Isaak}, {Marchenkov},
  {Miller}  \& {New}}{{Chaplin} et~al.}{2003}]{2003ESASP.517..247C}
{Chaplin} W.~J.,  {Elsworth} Y.,  {Isaak} G.~R.,  {Marchenkov} K.~I.,  {Miller}
  B.~A.,   {New} R.,  2003, in {Sawaya-Lacoste} H.,  ed.,  ESA Special
  Publication Vol. 517, GONG+ 2002. Local and Global Helioseismology: the
  Present and Future. pp 247--250

\bibitem[\protect\citeauthoryear{{Duvall}, {Harvey}, {Jefferies}  \&
  {Pomerantz}}{{Duvall} et~al.}{1991}]{1991ApJ...373..308D}
{Duvall} T.~L. J.,  {Harvey} J.~W.,  {Jefferies} S.~M.,   {Pomerantz} M.~A.,
  1991, \mn@doi [\apj] {10.1086/170052}, \href
  {https://ui.adsabs.harvard.edu/abs/1991ApJ...373..308D} {373, 308}

\bibitem[\protect\citeauthoryear{{Fossat} et~al.,}{{Fossat}
  et~al.}{1992}]{1992A&A...266..532F}
{Fossat} E.,  et~al., 1992, \aap, \href
  {https://ui.adsabs.harvard.edu/abs/1992A&A...266..532F} {266, 532}

\bibitem[\protect\citeauthoryear{{Fr{\"o}hlich} et~al.,}{{Fr{\"o}hlich}
  et~al.}{1995}]{1995SoPh..162..101F}
{Fr{\"o}hlich} C.,  et~al., 1995, \mn@doi [\solphys] {10.1007/BF00733428},
  \href {https://ui.adsabs.harvard.edu/abs/1995SoPh..162..101F} {162, 101}

\bibitem[\protect\citeauthoryear{{Fr{\"o}hlich} et~al.,}{{Fr{\"o}hlich}
  et~al.}{1997}]{1997SoPh..170....1F}
{Fr{\"o}hlich} C.,  et~al., 1997, \mn@doi [\solphys] {10.1023/A:1004969622753},
  \href {https://ui.adsabs.harvard.edu/abs/1997SoPh..170....1F} {170, 1}

\bibitem[\protect\citeauthoryear{{Gabriel} et~al.,}{{Gabriel}
  et~al.}{1995}]{1995SoPh..162...61G}
{Gabriel} A.~H.,  et~al., 1995, \mn@doi [\solphys] {10.1007/BF00733427}, \href
  {https://ui.adsabs.harvard.edu/abs/1995SoPh..162...61G} {162, 61}

\bibitem[\protect\citeauthoryear{{Garc{\'\i}a} et~al.,}{{Garc{\'\i}a}
  et~al.}{1998}]{1998ApJ...504L..51G}
{Garc{\'\i}a} R.~A.,  et~al., 1998, \mn@doi [\apjl] {10.1086/311561}, \href
  {https://ui.adsabs.harvard.edu/abs/1998ApJ...504L..51G} {504, L51}

\bibitem[\protect\citeauthoryear{{Garc{\'\i}a} et~al.,}{{Garc{\'\i}a}
  et~al.}{2005}]{2005A&A...442..385G}
{Garc{\'\i}a} R.~A.,  et~al., 2005, \mn@doi [\aap]
  {10.1051/0004-6361:20052779}, \href
  {https://ui.adsabs.harvard.edu/abs/2005A&A...442..385G} {442, 385}

\bibitem[\protect\citeauthoryear{Harris et~al.,}{Harris
  et~al.}{2020}]{harris2020array}
Harris C.~R.,  et~al., 2020, \mn@doi [Nature] {10.1038/s41586-020-2649-2}, 585,
  357

\bibitem[\protect\citeauthoryear{{Harvey}, {Tucker}  \& {Britanik}}{{Harvey}
  et~al.}{1998}]{1998ESASP.418..209H}
{Harvey} J.,  {Tucker} R.,   {Britanik} L.,  1998, in {Korzennik} S.,  ed.,
  ESA Special Publication Vol. 418, Structure and Dynamics of the Interior of
  the Sun and Sun-like Stars. p.~209

\bibitem[\protect\citeauthoryear{{Hill} et~al.,}{{Hill}
  et~al.}{1996}]{1996Sci...272.1292H}
{Hill} F.,  et~al., 1996, \mn@doi [Science] {10.1126/science.272.5266.1292},
  \href {https://ui.adsabs.harvard.edu/abs/1996Sci...272.1292H} {272, 1292}

\bibitem[\protect\citeauthoryear{{Howe}, {Chaplin}, {Davies}, {Elsworth},
  {Basu}  \& {Broomhall}}{{Howe} et~al.}{2018}]{2018MNRAS.480L..79H}
{Howe} R.,  {Chaplin} W.~J.,  {Davies} G.~R.,  {Elsworth} Y.,  {Basu} S.,
  {Broomhall} A.~M.,  2018, \mn@doi [\mnras] {10.1093/mnrasl/sly124}, \href
  {https://ui.adsabs.harvard.edu/abs/2018MNRAS.480L..79H} {480, L79}

\bibitem[\protect\citeauthoryear{Hunter}{Hunter}{2007}]{2007CSE.....9...90H}
Hunter J.~D.,  2007, \mn@doi [Computing in Science and Engineering]
  {10.1109/MCSE.2007.55}, 9, 90

\bibitem[\protect\citeauthoryear{{Jain} \& {Roberts}}{{Jain} \&
  {Roberts}}{1996}]{1996ApJ...456..399J}
{Jain} R.,  {Roberts} B.,  1996, \mn@doi [\apj] {10.1086/176662}, \href
  {https://ui.adsabs.harvard.edu/abs/1996ApJ...456..399J} {456, 399}

\bibitem[\protect\citeauthoryear{{Jain}, {Tripathy}, {Watson}, {Fletcher},
  {Jain}  \& {Hill}}{{Jain} et~al.}{2012}]{2012A&A...545A..73J}
{Jain} R.,  {Tripathy} S.~C.,  {Watson} F.~T.,  {Fletcher} L.,  {Jain} K.,
  {Hill} F.,  2012, \mn@doi [\aap] {10.1051/0004-6361/201219876}, \href
  {https://ui.adsabs.harvard.edu/abs/2012A&A...545A..73J} {545, A73}

\bibitem[\protect\citeauthoryear{{Jefferies}, {Pomerantz}, {Duvall}, {Harvey}
  \& {Jaksha}}{{Jefferies} et~al.}{1988}]{1988ESASP.286..279J}
{Jefferies} S.~M.,  {Pomerantz} M.~A.,  {Duvall} T.~L. J.,  {Harvey} J.~W.,
  {Jaksha} D.~B.,  1988, in {Rolfe} E.~J.,  ed.,  ESA Special Publication Vol.
  286, Seismology of the Sun and Sun-Like Stars. pp 279--284

\bibitem[\protect\citeauthoryear{{Jim{\'e}nez}}{{Jim{\'e}nez}}{2006}]{2006ApJ...646.1398J}
{Jim{\'e}nez} A.,  2006, \mn@doi [\apj] {10.1086/505105}, \href
  {https://ui.adsabs.harvard.edu/abs/2006ApJ...646.1398J} {646, 1398}

\bibitem[\protect\citeauthoryear{{Jimenez-Reyes}, {Regulo}, {Palle}  \& {Roca
  Cortes}}{{Jimenez-Reyes} et~al.}{1998}]{1998A&A...329.1119J}
{Jimenez-Reyes} S.~J.,  {Regulo} C.,  {Palle} P.~L.,   {Roca Cortes} T.,  1998,
  \aap, \href {https://ui.adsabs.harvard.edu/abs/1998A&A...329.1119J} {329,
  1119}

\bibitem[\protect\citeauthoryear{{Jim{\'e}nez}, {Roca Cort{\'e}s}  \&
  {Jim{\'e}nez-Reyes}}{{Jim{\'e}nez} et~al.}{2002}]{2002SoPh..209..247J}
{Jim{\'e}nez} A.,  {Roca Cort{\'e}s} T.,   {Jim{\'e}nez-Reyes} S.~J.,  2002,
  \mn@doi [\solphys] {10.1023/A:1021226503589}, \href
  {https://ui.adsabs.harvard.edu/abs/2002SoPh..209..247J} {209, 247}

\bibitem[\protect\citeauthoryear{{Jim{\'e}nez}, {Jim{\'e}nez-Reyes}  \&
  {Garc{\'{\i}}a}}{{Jim{\'e}nez} et~al.}{2005}]{2005ApJ...623.1215J}
{Jim{\'e}nez} A.,  {Jim{\'e}nez-Reyes} S.~J.,   {Garc{\'{\i}}a} R.~A.,  2005,
  \mn@doi [\apj] {10.1086/428879}, \href
  {http://adsabs.harvard.edu/abs/2005ApJ...623.1215J} {623, 1215}

\bibitem[\protect\citeauthoryear{{Jim{\'e}nez}, {Garc{\'\i}a}  \&
  {Pall{\'e}}}{{Jim{\'e}nez} et~al.}{2011}]{2011ApJ...743...99J}
{Jim{\'e}nez} A.,  {Garc{\'\i}a} R.~A.,   {Pall{\'e}} P.~L.,  2011, \mn@doi
  [\apj] {10.1088/0004-637X/743/2/99}, \href
  {https://ui.adsabs.harvard.edu/abs/2011ApJ...743...99J} {743, 99}

\bibitem[\protect\citeauthoryear{{Johnston}, {Roberts}  \& {Wright}}{{Johnston}
  et~al.}{1995}]{1995ASPC...76..264J}
{Johnston} A.,  {Roberts} B.,   {Wright} A.~N.,  1995, in {Ulrich} R.~K.,
  {Rhodes} E.~J. J.,   {D\"appen} W.,  eds,  Astronomical Society of the
  Pacific Conference Series Vol. 76, GONG 1994. Helio- and Astero-Seismology
  from the Earth and Space. p.~264

\bibitem[\protect\citeauthoryear{{Jones}}{{Jones}}{1989}]{1989SoPh..120..211J}
{Jones} H.~P.,  1989, \mn@doi [\solphys] {10.1007/BF00159876}, \href
  {https://ui.adsabs.harvard.edu/abs/1989SoPh..120..211J} {120, 211}

\bibitem[\protect\citeauthoryear{{Kiefer}, {Schad}, {Davies}  \&
  {Roth}}{{Kiefer} et~al.}{2017}]{2017A&A...598A..77K}
{Kiefer} R.,  {Schad} A.,  {Davies} G.,   {Roth} M.,  2017, \mn@doi [\aap]
  {10.1051/0004-6361/201628469}, \href
  {https://ui.adsabs.harvard.edu/abs/2017A&A...598A..77K} {598, A77}

\bibitem[\protect\citeauthoryear{{Kumar} \& {Lu}}{{Kumar} \&
  {Lu}}{1991}]{1991ApJ...375L..35K}
{Kumar} P.,  {Lu} E.,  1991, \mn@doi [\apjl] {10.1086/186082}, \href
  {https://ui.adsabs.harvard.edu/abs/1991ApJ...375L..35K} {375, L35}

\bibitem[\protect\citeauthoryear{{Kumar}, {Duvall}, {Harvey}, {Jefferies},
  {Pomerantz}  \& {Thompson}}{{Kumar} et~al.}{1990}]{1990LNP...367...87K}
{Kumar} P.,  {Duvall} T.~L. J.,  {Harvey} J.~W.,  {Jefferies} S.~M.,
  {Pomerantz} M.~A.,   {Thompson} M.~J.,  1990, {What are the Observed
  High-Frequency Solar Acoustic Modes?}.
p.~87, \mn@doi{10.1007/3-540-53091-6_68}

\bibitem[\protect\citeauthoryear{Lebigot}{Lebigot}{2019}]{uncertainties}
Lebigot E.~O.,  2019, {Uncertainties: a Python package for calculations with
  uncertainties}, \url {http://pythonhosted.org/uncertainties/}

\bibitem[\protect\citeauthoryear{{Libbrecht}}{{Libbrecht}}{1988}]{1988ApJ...334..510L}
{Libbrecht} K.~G.,  1988, \mn@doi [\apj] {10.1086/166855}, \href
  {https://ui.adsabs.harvard.edu/abs/1988ApJ...334..510L} {334, 510}

\bibitem[\protect\citeauthoryear{{Libbrecht} \& {Woodard}}{{Libbrecht} \&
  {Woodard}}{1990}]{1990Natur.345..779L}
{Libbrecht} K.~G.,  {Woodard} M.~F.,  1990, \mn@doi [\nat] {10.1038/345779a0},
  \href {https://ui.adsabs.harvard.edu/abs/1990Natur.345..779L} {345, 779}

\bibitem[\protect\citeauthoryear{{Rabello Soares}}{{Rabello
  Soares}}{2019}]{2019MNRAS.486.1847R}
{Rabello Soares} M.~C.,  2019, \mn@doi [\mnras] {10.1093/mnras/stz1005}, \href
  {https://ui.adsabs.harvard.edu/abs/2019MNRAS.486.1847R} {486, 1847}

\bibitem[\protect\citeauthoryear{Reback et~al.,}{Reback
  et~al.}{2019}]{Pandas2019}
Reback J.,  et~al., 2019, pandas-dev/pandas: v0.25.3,
  \mn@doi{10.5281/zenodo.3524604}

\bibitem[\protect\citeauthoryear{{R{\'e}gulo}, {Garc{\'\i}a}  \&
  {Ballot}}{{R{\'e}gulo} et~al.}{2016}]{2016A&A...589A.103R}
{R{\'e}gulo} C.,  {Garc{\'\i}a} R.~A.,   {Ballot} J.,  2016, \mn@doi [\aap]
  {10.1051/0004-6361/201425408}, \href
  {https://ui.adsabs.harvard.edu/abs/2016A&A...589A.103R} {589, A103}

\bibitem[\protect\citeauthoryear{{Rhodes} E.~J. et~al.,}{{Rhodes}
  et~al.}{2011}]{2011JPhCS.271a2029R}
{Rhodes} E.~J. J.,  et~al., 2011, in GONG-SoHO 24: A New Era of Seismology of
  the Sun and Solar-Like Stars. p. 012029,
  \mn@doi{10.1088/1742-6596/271/1/012029}

\bibitem[\protect\citeauthoryear{{Ronan}, {Cadora}  \& {Labonte}}{{Ronan}
  et~al.}{1994}]{1994SoPh..150..389R}
{Ronan} R.~S.,  {Cadora} K.,   {Labonte} B.~J.,  1994, \mn@doi [\solphys]
  {10.1007/BF00712900}, \href
  {https://ui.adsabs.harvard.edu/abs/1994SoPh..150..389R} {150, 389}

\bibitem[\protect\citeauthoryear{Seabold \& Perktold}{Seabold \&
  Perktold}{2010}]{seabold2010statsmodels}
Seabold S.,  Perktold J.,  2010, in 9th Python in Science Conference.

\bibitem[\protect\citeauthoryear{{Simoniello}, {Finsterle}, {Garc{\'\i}a},
  {Salabert}  \& {Jim{\'e}nez}}{{Simoniello}
  et~al.}{2009}]{2009AIPC.1170..566S}
{Simoniello} R.,  {Finsterle} W.,  {Garc{\'\i}a} R.~A.,  {Salabert} D.,
  {Jim{\'e}nez} A.,  2009, in {Guzik} J.~A.,  {Bradley} P.~A.,  eds,  American
  Institute of Physics Conference Series Vol. 1170, American Institute of
  Physics Conference Series. pp 566--568, \mn@doi{10.1063/1.3246563}

\bibitem[\protect\citeauthoryear{{Tapping}}{{Tapping}}{2013}]{2013SpWea..11..394T}
{Tapping} K.~F.,  2013, \mn@doi [Space Weather] {10.1002/swe.20064}, \href
  {https://ui.adsabs.harvard.edu/abs/2013SpWea..11..394T} {11, 394}

\bibitem[\protect\citeauthoryear{{Tripathy}, {Kumar}, {Jain}  \&
  {Bhatnagar}}{{Tripathy} et~al.}{2001}]{2001SoPh..200....3T}
{Tripathy} S.~C.,  {Kumar} B.,  {Jain} K.,   {Bhatnagar} A.,  2001, \mn@doi
  [\solphys] {10.1023/A:1010318428454}, \href
  {https://ui.adsabs.harvard.edu/abs/2001SoPh..200....3T} {200, 3}

\bibitem[\protect\citeauthoryear{Virtanen et~al.,}{Virtanen
  et~al.}{2020}]{2020SciPy-NMeth}
Virtanen P.,  et~al., 2020, \mn@doi [Nature Methods]
  {10.1038/s41592-019-0686-2}, 17, 261

\bibitem[\protect\citeauthoryear{{Vorontsov}, {Jefferies}, {Duval}  \&
  {Harvey}}{{Vorontsov} et~al.}{1998}]{1998MNRAS.298..464V}
{Vorontsov} S.~V.,  {Jefferies} S.~M.,  {Duval} T.~L. J.,   {Harvey} J.~W.,
  1998, \mn@doi [\mnras] {10.1046/j.1365-8711.1998.01630.x}, \href
  {https://ui.adsabs.harvard.edu/abs/1998MNRAS.298..464V} {298, 464}

\makeatother
\end{thebibliography}

\appendix
\section{GOLF supplemental information}
\begin{table*}
	\caption{Changes made to the GOLF detectors and their mode of operation indicated in Figure~\ref{fig:3}.$^{a}$}\label{tab:A1}
\begin{center}
\begin{tabular}{|c|c|}
	\hline
	\rule[-1ex]{0pt}{2.5ex} Date & Event \\
	\hline\hline
	\rule[-1ex]{0pt}{2.5ex} 16 February 1996 & Increase of voltage applied to the detectors$^b$  \\
	\rule[-1ex]{0pt}{2.5ex} 16 March 1996 & Increase of voltage applied to the detectors$^b$   \\
	\rule[-1ex]{0pt}{2.5ex} 17 October 1997 & Increase of voltage applied to the detectors  \\
	\rule[-1ex]{0pt}{2.5ex} 25 June 1998 & Switch from blue wing mode to red wing mode\\
	\rule[-1ex]{0pt}{2.5ex} 12 May 2000 & Increase of voltage applied to the detectors  \\
	\rule[-1ex]{0pt}{2.5ex} 18 November 2002 & Switch from red wing mode to blue wing mode\\
	\rule[-1ex]{0pt}{2.5ex} 22 April 2005 & Increase of voltage applied to the detectors  \\
	\rule[-1ex]{0pt}{2.5ex} 06 March 2014 & Increase of voltage applied to the detectors  \\
	\hline
	\multicolumn{2}{l}{$^a$ The dates and events are documented in more detail on the instrument website: \url{https://www.ias.u-psud.fr/golf/templates/index.html}}\\
	\multicolumn{2}{l}{$^b$ Not indicated in Figure~\ref{fig:3}.}\\
\end{tabular}
\end{center}
\end{table*}
\section{Additional plots}
\begin{figure*}
    \centering
    \includegraphics[width=0.8\linewidth]{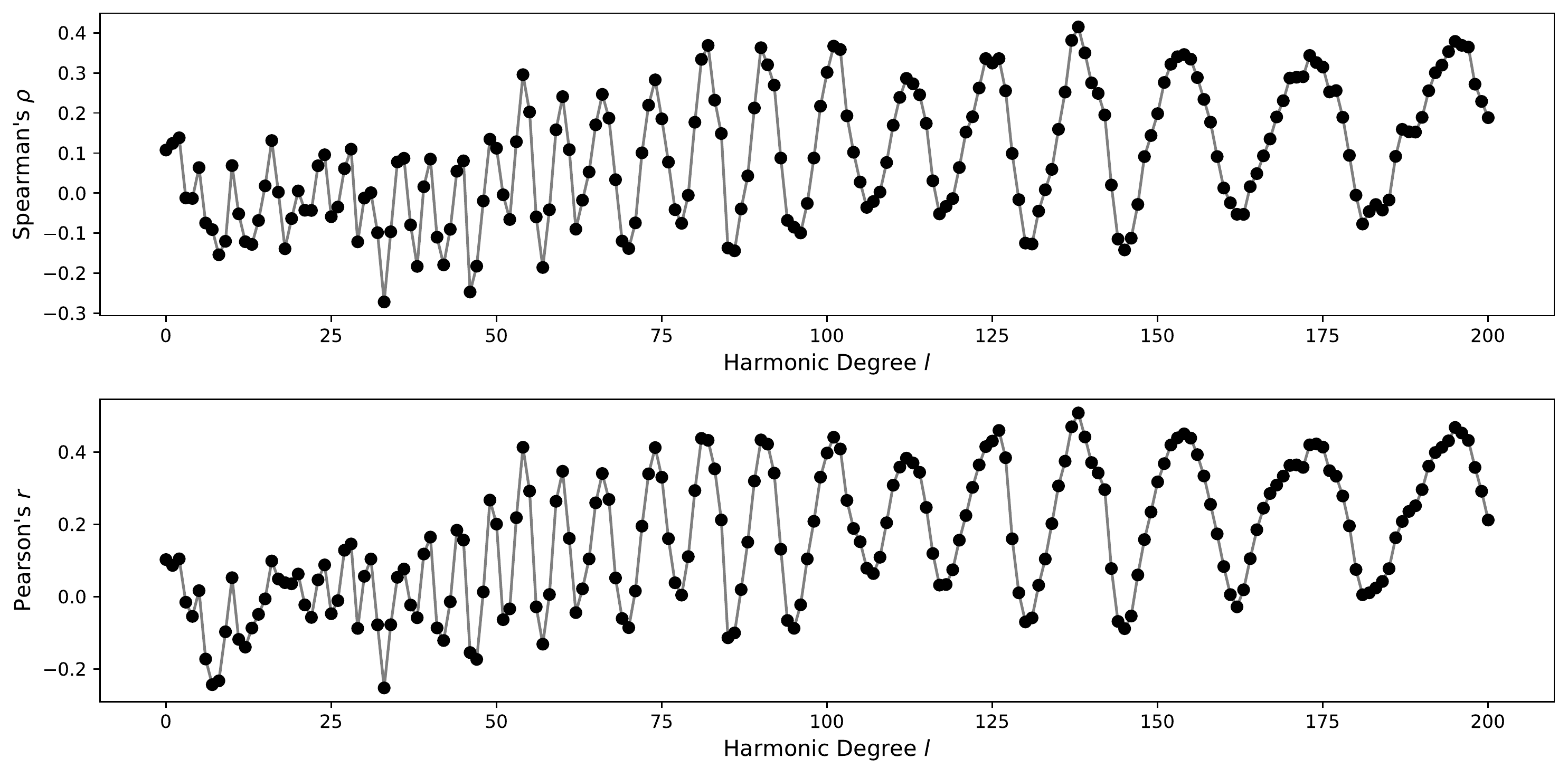}
    \caption{Top panel: Spearman rank correlation $\rho$ between pseudomode frequency shift and the fill factor of the GONG segments over all $l$. Bottom panel: Pearson correlation coefficient $r$ between pseudomode frequency shift and the fill factor of the GONG segments over all $l$.} 
    \label{app:fig:1}
    \includegraphics[width=0.8\linewidth]{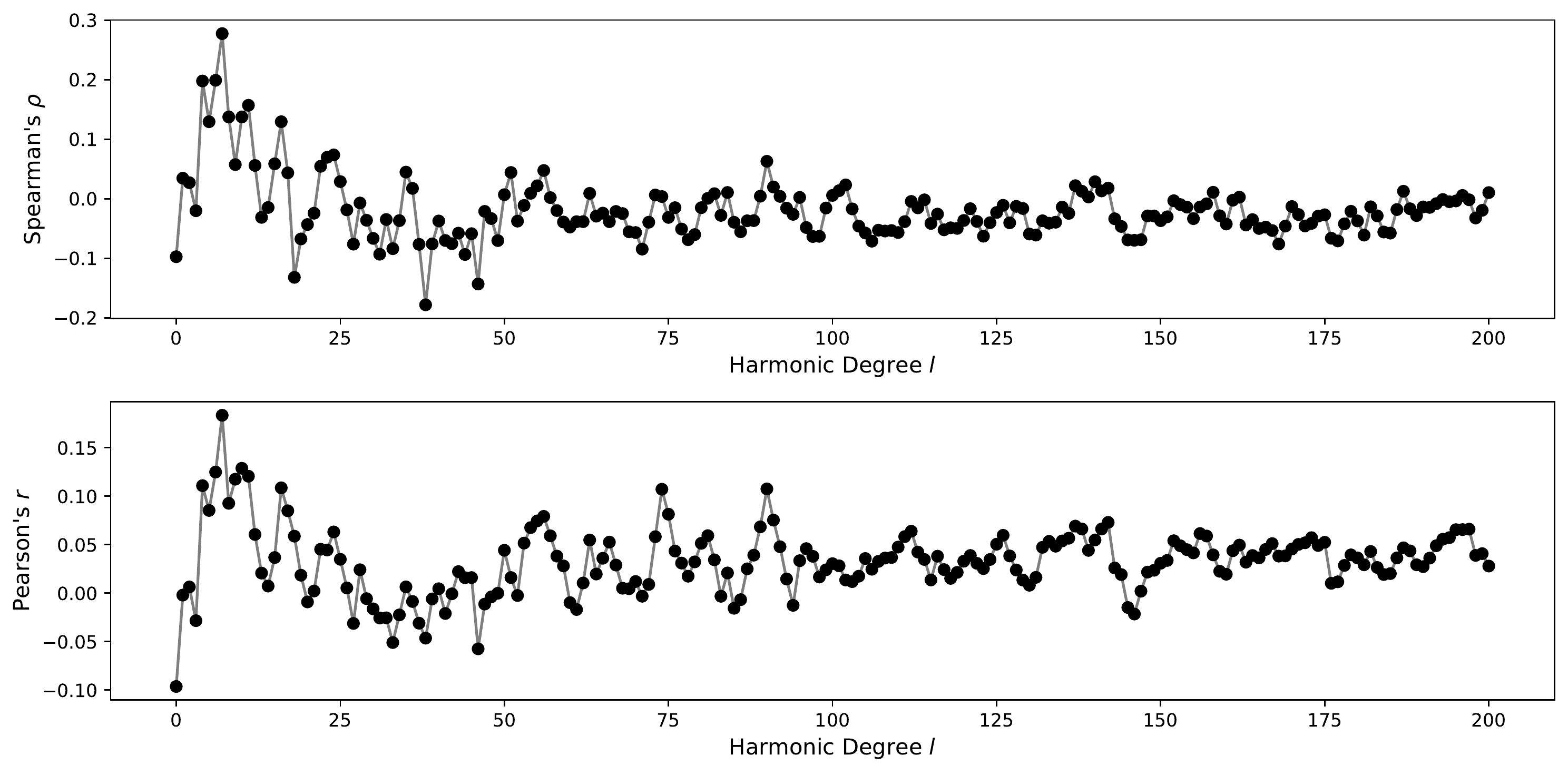}
    \caption{Same as Fig.~\ref{app:fig:1} but after the linear fill correction to the shifts described in Section~\ref{sect:method} was applied.} 
    \label{app:fig:2}
\end{figure*}


\bsp	
\label{lastpage}
\end{document}